\documentclass[aps,pra,reprint,showpacs,twocolumn,superscriptaddress,eqsecnum]{revtex4-1}
\usepackage{graphicx}
\usepackage{amsmath}
\usepackage{amssymb}
\usepackage{amsfonts}
\usepackage{bbm}
\usepackage{braket}
\usepackage{bm}
\usepackage{verbatim}
\usepackage{cancel}
\usepackage{tikz}
\usepackage{wasysym}
\usepackage[bookmarks=false,colorlinks=true,urlcolor=blue,citecolor=blue,linkcolor=blue]{hyperref}
\usepackage[normalem]{ulem}
\allowdisplaybreaks

\definecolor{darkblue}{RGB}{0,0,127}

\newcommand{\dexpect}[1]{\left\langle\kern -2pt \left\langle{#1}\right\rangle\kern -2pt \right\rangle}
\graphicspath{{figures/}}

\begin{document}

\title{Phase transitions in the presence of fluctuating charge-density wave in two-dimensional film of kagome metals}

\author{Julia Wildeboer}
\affiliation{Division of Condensed Matter Physics and Materials Science, Brookhaven National Laboratory, Upton, NY 11973-5000, USA}

\author{Saheli Sarkar}
\affiliation{Division of Condensed Matter Physics and Materials Science, Brookhaven National Laboratory, Upton, NY 11973-5000, USA}
\affiliation{Harish-Chandra Research Institute, A CI of Homi Bhabha National Institute, Chhatnag Road, Jhunsi, Prayagraj - 211019, India}

\author{Alexei M. Tsvelik}
\affiliation{Division of Condensed Matter Physics and Materials Science, Brookhaven National Laboratory, Upton, NY 11973-5000, USA}

\date{\today}

\begin{abstract}
We determine the nature of a phase transition in a model describing an interaction of multiple charge density waves in a two dimensional film. The model was introduced by two of the authors 
in Phys. Rev. B {\bf 108}, 045119 (2023) to describe fluctuations in charge density wave order in the kagome metals AV$_3$Sb$_5$ (A=K, Rb, Cs) in two dimensions. The situation is nontrivial since the transition occurs in the region of phase diagram where the unbound vortices compete with the interaction between charge density waves.  
Here, we study the nature of the phase transition via Metropolis Monte Carlo simulations. The 3-component order parameter, the  susceptibility, the energy per site, and
the specific heat are measured for a range of temperatures for different lattice sizes $L=8,16,24,32$. The finite size scaling 
analysis indicates the presence of a second-order transition. 
\end{abstract}


\maketitle
\section{Introduction}
Two-dimensional (2D) correlated metals derived from transition metal ions represent a significant area of study within many-body physics~\cite{kennes2021moire,dzero2016topological}. These materials exhibit a rich range of intriguing electronic phenomena, including unconventional superconductivity~\cite{Gomes,Teicher,Sarte}, charge and spin ordering~\cite{jiang2021unconventional,Li}, nematicity~\cite{uykur2022optical,Ruff}, and strange metallic behavior~\cite{Tan,Denner,feng2021chiral}. While the majority of theoretical and experimental investigations have focused on structures with square lattices, such as cuprates~\cite{Teicher,yu2107evidence} and iron-based superconductors~\cite{Gomes,yang2020giant,YuWu}, there is comparatively less research on correlated metals with hexagonal and triangular symmetries.  
Recently, a new class of kagome metals has emerged, offering exciting opportunities for further exploration. 
The  tight-binding electronic structure of the kagome metals~\cite{yin2022topological} feature a flat band, Dirac points and van-Hove singularities which naturally leads to the existence of various types of order which precise nature depends on the filling fractions of electrons in the band and the nature of electronic interactions.  
Excellent examples of such systems are recently discovered quasi-two dimensional kagome lattice metals AV$_3$Sb$_5$ (A=K, Rb, Cs)~\cite{OrtizB2019}. The electron filling in these metals is such that the chemical potential is very close to the van-Hove singularities. The different types of order are likely to originate from the  electron-electron interactions at the nested Fermi surface.
The observed phases include charge-density wave (CDW) order~\cite{jiang2021unconventional}, superconductivity ~\cite{ortiz2019new} and pair-density wave ~\cite{chen2021roton,PDWLiPRX2023} among others. 
These experimental observations have inspired a search for something more exotic. As we have demonstrated in our previous work ~\cite{Sarkar2023}, a competition between order parameter fluctuations and topological excitations in 2D films can lead to different composite orders. 
 
\begin{figure}
 \includegraphics[width=\columnwidth]{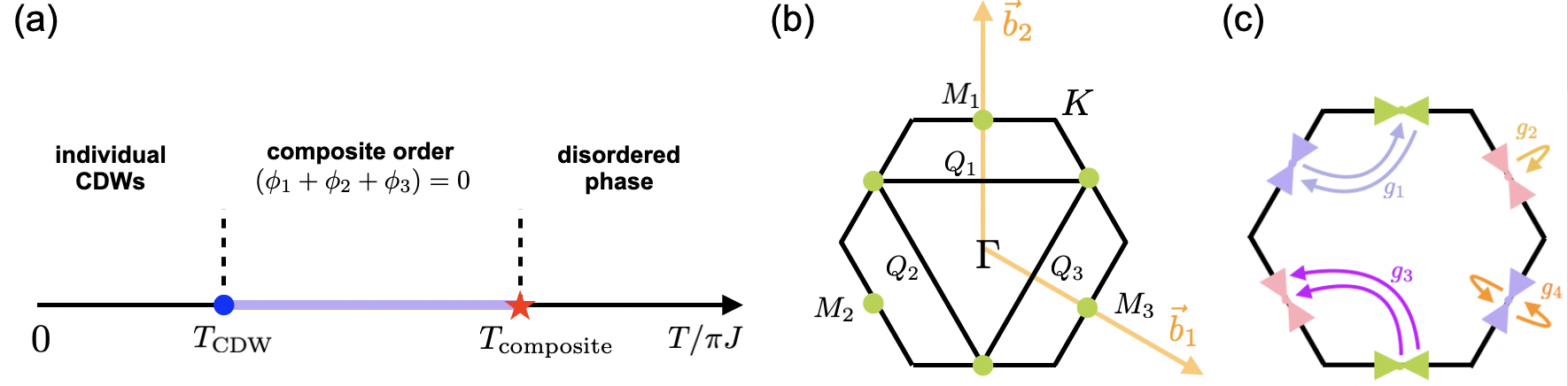}
 \caption{
        (a) A schematic phase diagram of the model described by the free energy functional Eqn.~\eqref{GL}, [adopted from \cite{Sarkar2023}]. For $J=1$, there is a crossover into a regime with composite order around $T_{\rm composite}$, where the sum of the charge-density wave (CDW) phases are frozen, i.e. $(\phi_{1}+\phi_{2}+\phi_{3}) = 0$. 
        Then at a further low temperature $T_{\rm CDW}$, there is a phase transition into the state where individual CDWs order. For $g_{3} > 0$, the low-temperature phase breaks time-reversal symmetry. 
       (b) The hexagonal Brillouin zone (BZ) of the kagome lattice, with reciprocal lattice vectors $\vec{b}_{1} = 4\pi(0, 1/\sqrt{3})$ and $\vec{b}_{2} = 2\pi (1,-1/\sqrt{3})$. The BZ also portrays the center (${\bf \Gamma}$) at (0,0) and the high symmetry $K$ points located at the vertices of the BZ and the 
       points $\bf{M}_{a}$ with $a=1,2,3$ at positions ${\bf M}_1= \frac{1}{2}\vec{b_2}$, ${\bf M}_2= -\frac{1}{2}(\vec{b_1} + \vec{b_2})$ and ${\bf M}_3= \frac{1}{2}\vec {b_1}$ located at the hexagonal Brillouin zone face centers.  
       Compounds such as AV$_3$Sb$_5$ with A = K, Rb, Cs exhibit saddle points at the $\vec{M_{a}}$, $a=1,2,3$, shown by green circles. The $M_{a}$ points are connected by the three nesting vectors ${\bf Q}_{a}$, $a=1,2,3$ that also serve as  the ordering wave-vectors of the CDW.
       (c) Depicted are all possible interactions $g_{1}$, $g_{2}$, $g_{3}$ and $g_{4}$ in the patch model (see Hamiltonian~\eqref{H1}). The cones represent the saddle-point dispersion at each ${\bf M}_{a}$ while the arrows depict the scattering processes described by the interactions. 
}
\label{fig:CDW}
\end{figure}

In this work we concentrate our attention on the charge density wave state. CDW order in kagome materials AV$_3$Sb$_5$ (A=K, Rb, Cs) is quite complex since it has three CDWs interacting with each other (see Fig.~\ref{fig:CDW}) and it breaks time reversal  ~\cite{TRS_ThomalePRL2021,mielke2022time} and six-fold rotational symmetries \cite{li2022rotation}. 
In three dimensional materials, where the coupling between the layers is sufficiently strong~\cite{barman2024stacking}, the phase transition to the CDW state when all three wave components order is of the second order. In 2D films ~\cite{kim2023monolayer,Songprl2021,wang2021enhancement,thinZhangPRB2022,zheng2023electrically} there are strong phase fluctuations ~\cite{WangPRBfluc2014,sarkarloopcurrentprb2019} and there are also topologically nontrivial  configurations, i.e. vortices which influence the ordering~\cite{sachdev2018topological}. 

\textcolor{black}{In our previous work, Ref.~\cite{Sarkar2023}, we demonstrated that the competition among CDW order parameters, fluctuations, and topological vortices leads to the condensation of a global phase prior to the emergence of individual CDW orders. In that study, we derived a Ginzburg-Landau functional, showing that the nature of the phase transition is governed by the interplay of two primary operators: one arising from Umklapp scattering and the other from vortex configurations.
This competition makes the problem especially interesting giving rise to a non-trivial situation, even in the case of a phase transition of these fluctuating CDW orders.
In this work, we address the open question of the character of the phase transition by employing Monte Carlo simulations utilizing the Ginzburg-Landau (GL) free energy functional derived in Ref.~\cite{Sarkar2023}. This functional describes the thermodynamic properties of the CDW state in 2D films of the kagome materials AV$_3$Sb$_5$ (A = K, Rb, Cs), incorporating vortex contributions. We note that a similar model applies to interacting superconducting order parameters in three-band $s$-wave superconductors (see, for instance, Refs.\cite{volkov, babaev}).
}
\begin{figure*}
    \includegraphics[width=1.00\textwidth]{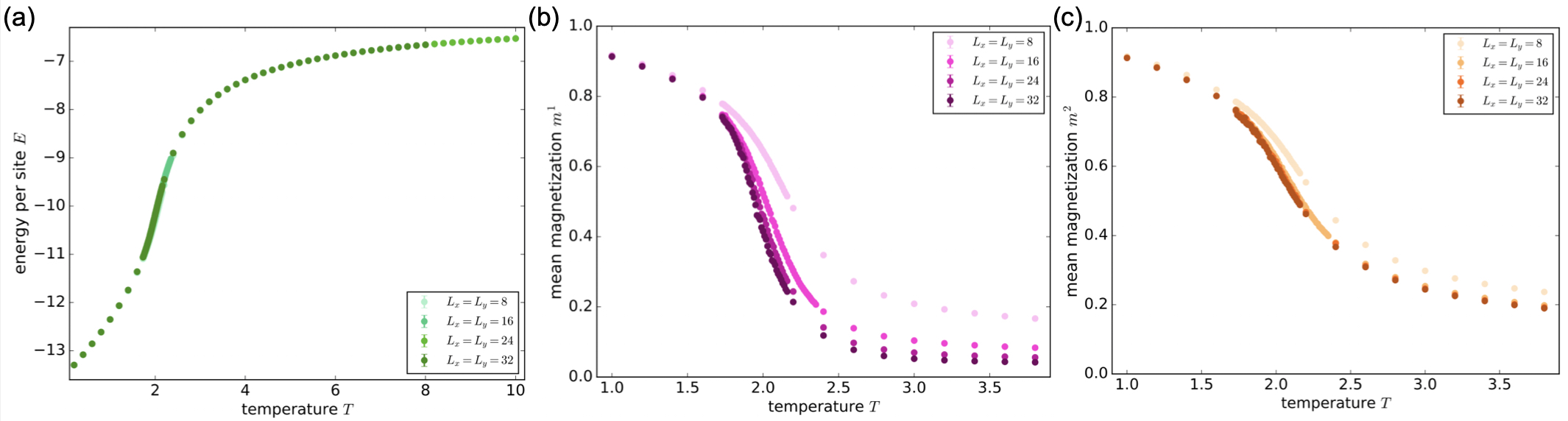}
\caption{
   (a) Energy per site $E$ as a function of temperature $T$ for the model described in Eq.~\eqref{ham}, with system parameters set to $(G, g_{3}) = (6.0, 1.0)$ and $(J, \Delta) = (1.0, 1.0)$. The plots correspond to system sizes of $L_{x} = L_{y} \equiv L = 8, 16, 24, 32$. As the system undergoes a second-order phase transition at temperature $T_c$, the energy changes continuously, exhibiting no discontinuities. 
   (b) The mean magnetization $m^{1}$ versus temperature $T$ illustrates the behavior of the order parameter as the system approaches the critical temperature $T_c$. The continuous nature of the order parameter indicates a second-order phase transition. Notably, the value of $m^{1}$ above $T_c$ is influenced by the lattice size $L$, with the continuum limit $\lim_{T \to \infty} m^{i} \rightarrow 0$ approached more closely as $L$ increases. 
   (c) The second order parameter $m^{2}$ exhibits analogous behavior to $m^{1}$, while the third order parameter $m^{3}$ (not shown) follows a similar trend. Together, these three order parameters underscore the continuous transition characteristics inherent in second-order phase transitions. 
}\label{fig2}
\end{figure*}
\section{The patch model}
To ensure a comprehensive understanding, we will revisit some analyses from our previous work~\cite{Sarkar2023}, where we investigated the multi-component CDW order beyond mean-field theory. The key outcome of that study was the derivation of the Ginzburg-Landau functional, which incorporated both phase fluctuations of the CDW order parameters and their topologically nontrivial configurations. The derivation relied on the so-called patch model~\cite{Park}, which specifically examines the Vanadium atoms in AV$_3$Sb$_5$ and predicts the formation of van Hove singularities (VHSs) at the three 
${\bf M}_{a}$-points, $i=1,2,3$, within the Brillouin zone. 

The patch model Hamiltonian consists of two parts, i.e. $\mathcal{H}_{\rm patch} = \mathcal{H}_{0} + \mathcal{H}_{1}$. The non-interacting part $\mathcal{H}_{0}$ describes a low-energy continuum model by taking patches around the ${\bf M}_{a}$-points in the Brillouin zone with a cutoff radius $\Lambda$: 
\begin{eqnarray}\label{H0}
\mathcal{H}_{0} = \sum_{a=1}^{3}\sum_{|{\bf k}|< \Lambda} c_{a {\bf k}}^{\dagger}\left[\epsilon_{a}({\bf k}))-\mu\right] c_{a {\bf k}} 
\end{eqnarray} 
with $a$ being the patch index while ${\bf k}$ is the momentum measured from ${\bf M_{a}}$, $\epsilon_{a}({\bf k})$ is the saddle-point dispersion sitting at ${\bf M_{a}}$, and $\mu$ is the chemical potential measured away from the saddle point. 
The single electron saddle point dispersions take the form 
\begin{eqnarray}\label{disper}
\epsilon_{1} = k_{1} (k_{1} + k_{2}) && \epsilon_{1} = - k_{1} k_{2} \nonumber \\
\epsilon_{3} = k_{2} (k_{1} + k_{2}) &&  k_{1,2} = k_{x} \pm \sqrt{3} k_{y}
\end{eqnarray}
The four possible electron-electron interactions between the three patches are captured by 
\begin{eqnarray}\label{H1}
\mathcal{H}_{1} &=& \sum_{{\bf k_1}, {\bf k_2}, {\bf k_3}, {\bf k_4}} \Bigg[\sum_{a \neq b} \Big( 
g_{1}c_{a {\bf k_{1}} \sigma}^{\dagger}c_{b {\bf k_{2}} \sigma^{\prime}}^{\dagger}c_{a {\bf k_{3}} \sigma^{\prime}}c_{b {\bf k_{4}} \sigma} \nonumber \\
&+&g_{2}c_{a {\bf k_{1}} \sigma}^{\dagger}c_{b {\bf k_{2}} \sigma^{\prime}}^{\dagger}c_{a {\bf k_{3}} \sigma^{\prime}}c_{b {\bf k_{4}} \sigma} \nonumber \\
&+&g_{3}c_{a {\bf k_{1}} \sigma}^{\dagger}c_{a {\bf k_{2}} -\sigma^{\prime}}^{\dagger}c_{b {\bf k_{3}} -\sigma^{\prime}}c_{b {\bf k_{4}} \sigma}\nonumber \\
&+&\sum_{a} g_{4}c_{a {\bf k_{1}} \sigma}^{\dagger}c_{a {\bf k_{2}} -\sigma^{\prime}}^{\dagger}c_{b {\bf k_{3}} -\sigma^{\prime}}c_{b {\bf k_{4}} \sigma}
\Big) \Bigg]
\end{eqnarray} 
In the above Eqn.~\eqref{H1}, we
have the constraint ${\bf k_{1}} + {\bf k_{2}} + {\bf k_{3}} + {\bf k_{4}} = 0$. 
Moreover, the $g_1$, $g_2$, $g_3$, $g_4$ coupling parameters represent interpatch exchange, interpatch density-density, Umklapp, and intrapatch density-density scattering processes, respectively (see Fig.~\ref{fig:CDW}(c)). 

\textcolor{black}{The analysis of model~\eqref{H1} conducted in previous papers (see~\cite{Sarkar2023} and references therein) indicates a possibility of CDW order. The corresponding CDW order parameter fields are 
\begin{eqnarray}
\Delta_{ij}(x) = \sum_{\sigma}c^+_{k_i\sigma}(x)c_{k_j\sigma}(x) 
\end{eqnarray}
where the $k$'s are located at the saddle points of the Brillouin zone. The corresponding} GL functional for the free energy ~\cite{Sarkar2023}, includes three phase fields $\phi_{a}$ of the corresponding CDW order parameter fields given by $\Delta_{a} = |\Delta_a|\exp[i\phi_a]$, with 
$a=1, 2, 3$ 
for the three CDWs. The wave-vectors ${\bf Q}$s of these three CDWs are shown in the Figure~\ref{fig:CDW}(b). 

The lattice version of the GL functional is given by 
\begin{eqnarray}\label{GL}
{\cal F} /T &=& -\frac{1}{T}\Big\{J\sum_{a=1,2,3} \sum_{<b>}\frac{1}{b^2}\cos\Big[\phi_a({\bf x}) - \phi_a({\bf x} + {\bf b})\Big]\nonumber \\
&+& G\cos[\phi_1({\bf x}) +\phi_2({\bf x}) +\phi_3({\bf x})]\nonumber \\ 
&+&  g_3\Delta^2\sum_{a=1,2,3}\cos[2\phi_a({\bf x})]\Big\}\,.
\end{eqnarray} 
Here, the parameter $J$ is the phase stiffness while $T$ is the temperature. The $g_3$-term is due to the Umklapp scattering, the $G$-term describes the coupling between the CDWs. Only the coupling of all three is allowed by the quasi momentum conservation, i.e. ${\bf Q}_1 +{\bf Q}_2 + {\bf Q}_3 = 0$.  
The lattice regularization properly takes into account the vortices whose presence is hidden in the periodicity of the cosine function. 
Close to the phase transition, the precise lattice geometry is irrelevant. 
Therefore, in our numerical analysis we will consider the somewhat simpler geometry of the square lattice with lattice constant $b$. 
In two dimensions, it is possible for the phases of three CDWs to lock while the individual CDW orders do not form. To illustrate this, we consider the case of large $G$, where the sum of all three phases is constrained to zero.
Now we can make a transformation and introduce new fields as follows 
\begin{eqnarray}
\phi_a = \frac{\Phi}{\sqrt{3}} + \sqrt{\frac{2}{3}}~
({\bf e}_a {\bm{\varphi}}) 
\end{eqnarray}
with 
\begin{eqnarray}
{\bf e}_a = (1,0), ~(-1/2, \sqrt 3/2), ~(-1/2,-\sqrt 3/2) 
\end{eqnarray}
and ${\bm \varphi} = (\varphi_1,\varphi_2)$, $a = 1, 2, 3$ and treat $\Phi$ as gapped. To take  into account the vortex configurations we also need to introduce dual fields $\vartheta_j$. Following standard procedures we  arrive  at the continuum version of the model (\ref{GL})~\cite{Sarkar2023}: 
 \begin{eqnarray}
 \label{GL2}
 &&{\cal F}_{eff}/T =\nonumber \\
 && \int d^2x \Big\{\sum_{a=1}^3\Big[ \bar g_3 \cos(\sqrt{8/3}{\vec e}_a\vec\varphi) - B\cos(2\pi\sqrt{2}\vec\omega_a{\vec\vartheta}) \Big]+ \nonumber\\
 &&\sum_{j=1,2}\Big[\frac{1}{2T}(\partial_x\varphi_j)^2 +\frac{T}{2}(\partial_x\vartheta_j)^2 +i \partial_x\varphi_j\partial_y\vartheta_j\Big]\Big\}, 
\end{eqnarray}
where  $\vec\omega_a = (0,1), (\sqrt 3, 1)/2, (\sqrt 3 ,-1)/2$ and $\bar g_3,B$ are coupling constants related to the original ones, we set $J=1$.

The distinct feature of action~\eqref{GL2} is that it contains competing cosine terms. The ones with $\varphi_a$ fields are relevant at small $T$ such that individual CDWs acquire finite average amplitudes. The dual cosines corresponding to vortex configurations (the ones with disorder fields $\vartheta$) dominate at higher temperatures. The transition occurs in the temperature interval where both types of operators are relevant and is determined by their competition. 

The model~\eqref{GL2} and its lattice regularization~\eqref{GL} belong to the class of affine XY models. Some of them  have been studied in relation to the problem of quark confinement ~\cite{anber20122d,anber20133}. The numerics~\cite{anber20133} indicate that the phase transition may be a weak first order transition.  
In the following we employ the Metropolis Monte Carlo method similar to the one used in~\cite{anber20133} for model~\eqref{GL}. 
\begin{figure*}
    \includegraphics[width=1.0\textwidth]{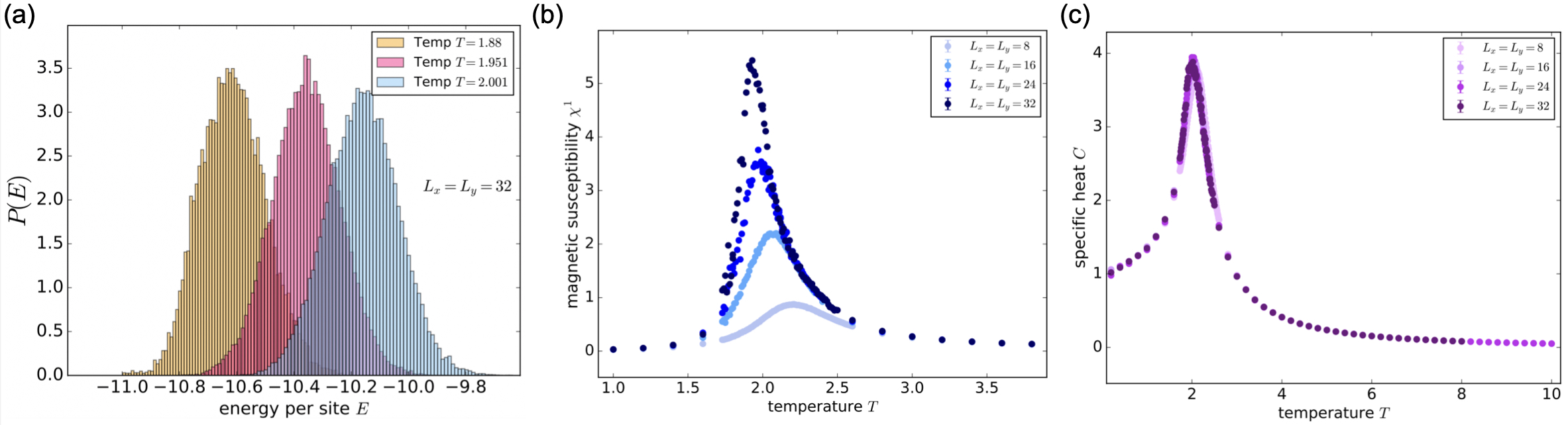}
\caption{
    (a) Probability distribution $P(E)$ of the energy per site $E$ for a square lattice of size $L = 32$ at three distinct temperatures: $T = 1.88$ (below the critical temperature $T_c$), $T = 1.951$ (approximately $T_c$), and $T = 2.001$ (above $T_c$). The parameters are set to $(G, g_{3}) = (6.0, 1.0)$ and $(J, \Delta) = (1.0, 1.0)$. The energy histograms reveal that there is no double peak distribution across the observed temperature range, consistent with the characteristics of a second-order phase transition. The absence of a well-defined double peak is evident, and all histograms maintain a similar shape throughout the temperature interval $T \in [1.88, 2.118]$ with increments of $\Delta T = 0.001$.
   (b) The magnetic susceptibility $\chi^{1}$ displays a trend of being approximately zero at both high and low temperatures, with a notable peak emerging around the critical temperature $T_c$. The peak height is influenced by the lattice size $L$. The second component of susceptibility, $\chi^{2}$, is presented in the inset and exhibits qualitatively similar behavior, as does the third component (not shown).
   (c) The specific heat $C$ as a function of temperature $T$ transitions from a value of $C = 1$ at low temperatures to a sharp peak near the critical temperature, followed by a decrease at higher temperatures. The peak height varies with the size of the lattice, reinforcing the effects of the phase transition on the specific heat capacity. 
}
    \label{fig3}
\end{figure*}
\section{Monte Carlo Simulations} 
In this section, we investigate the effects of varying temperature $T$ and coupling constants $G$ and $g_{3}$ in the Ginzburg-Landau (GL) functional~\eqref{GL}, setting the parameters $J$, $b$, and $\Delta$ to one (i.e., $J = b = \Delta = 1$). The free energy expressed in~\eqref{GL} incorporates the dynamics of mean-field fluctuations, thus serving as the Hamiltonian necessary for our Monte Carlo (MC) simulations. 

This leads to the formulation of the Boltzmann factor, i.e. $e^{-\mathcal{F}/T}$, which is evaluated at every iteration of the Metropolis algorithm employed in our MC simulations. To facilitate the computation of the average energy per site later, we also present the Hamiltonian that describes the system as follows~\footnote{Instead of simulating the Boltzmann factor $e^{-\mathcal{F}/T}$, one can equivalently evaluate the factor $e^{-\mathcal{H}/T}$ from Eqs.~\eqref{ham}.} 
\begin{eqnarray}\label{ham}
\mathcal{H} &=& -J\sum_{{\bf r}}\sum_{a=1,2,3} \sum_{<b>}\frac{1}{b^2}\cos\Big[\phi_a({\bf r}) - \phi_a({\bf r} + {\bf b})\Big] \nonumber \\
&-& G\sum_{{\bf r}}\cos(\phi_{1}({\bf r}) +\phi_{2}({\bf r}) + \phi_{3}({\bf r})) \\
&+& \Delta^{2} g_{3}\sum_{{\bf r}}\sum_{a=1,2,3}\cos[2\phi_a({\bf r})]\nonumber\,
\end{eqnarray}
with $\phi_{a} \in [0; 2\pi]$ with $a=1,2,3$ and ${\bf r} = (x,y)$ runs over all lattice sites. 
We perform simulations on a square lattice of size $(L_{x}, L_{y})$  with $L_{x} = L_{y} \equiv L$, leading to a total of $N = L^{2}$  lattice sites. The lattice sizes examined include $L = 8, 16, 24, 32$, with varying the ratio of the parameters $(G, g_{3})$ using the Metropolis algorithm. The data is collected after $30,000$ Monte Carlo sweeps, with each sweep comprising $N$ Metropolis iterations. Measurements are taken every $10$ sweeps, and the first $3,000$ sweeps are discarded for equilibration purposes. 

Annealing is initiated from a random configuration at high temperatures to ensure effective data collection. In the critical region, where phase transitions occur, the temperature intervals $\Delta T$ are reduced to $0.01$, and further refined to $\Delta T = 0.001$ during finite-size scaling analyses.

We define the quantities measured in our simulations as follows. The average energy per site is given by: 
\begin{eqnarray}\label{specific_heat}
E = \frac{\langle \mathcal{H} \rangle}{N}\,.
\end{eqnarray}
The ``magnetization'' $M^{j}$ for each field component $(j = 1, 2, 3)$ is defined as the average $j$-th ``spin,'' represented as a planar unit vector $e^{i\phi^{j}_{x}}$ over the entire lattice:
\begin{eqnarray}\label{magnetization}
M^{j} = \sum_{x} e^{i\phi^{j}_{x}}\,\,\,\,\,\,\text{with}\,\, j = 1, 2, 3\,.
\end{eqnarray}
The order parameter used to detect phase transitions is the mean magnetization $m^{j}$ per site, defined as:
\begin{eqnarray}\label{o_parameter}
m^{j} = \frac{1}{N}\langle |\sum_{x} e^{i\phi^{j}_{x}}| \rangle = \frac{\langle |M^{j}| \rangle}{N}\,\,\,\,\,\,\text{with}\,\, j = 1, 2, 3\,.
\end{eqnarray}
Additionally, we specify the magnetic susceptibility $\chi^{j}$ of each component as follows: 
\begin{eqnarray}\label{mag_sus}
\chi^{j} = \frac{\langle |M^{j}|^{2} \rangle - \langle |M^{j}| \rangle^{2}}{N T}\,\,\,\,\,\,\text{with}\,\, j = 1, 2, 3\,.
\end{eqnarray}
Lastly, the specific heat is calculated as: 
\begin{eqnarray}\label{specific_heat}
C = \frac{\langle \mathcal{H}^{2} \rangle - \langle \mathcal{H} \rangle^{2}}{N T^{2}}\,.
\end{eqnarray}
\subsection{MC results: Energy and magnetization}
Our goal is to determine the temperature $T_{c}$ at which a phase transition takes place. Further we want to identify the nature of the phase transition, i.e. whether it is a first or second order transition. 
We start by analyzing systems of varying size $L = 8, 16, 24, 32$ for coupling constants fixed $(G,g_{3}) = (6.0,1.0)$. 
Each type of the transition has its characteristic behavior. 
In case of a first order transition the energy per site typically shows a discontinuous jump at the transition temperature $T_{c}$,  indicating a latent heat. 
For our working model Eqn.~\eqref{ham}, the energy per site $E$ with temperature $T$ is plotted in Figure~\ref{fig2}(a) for various system sizes $L$ with couplings constants $(G,g_{3}) = (6.0,1.0)$. We do not observe any change in the $E$ with increase in system size. Importantly, there is {\it no discontinuity} in the energy per site. 

The local order parameter of the system $m^{j}$, $j=1,2,3$, is shown in Figure~\ref{fig2}(b) and (c) as a function of the temperature $T$ for lattices of width $L = 8, 16, 24, 32$. It is obvious that the magnetizations $m^{1}$, $m^{2}$ and $m^{3}$ behave identically. 
In all cases, the magnetization $m^{j}$ at higher temperatures above the observed transition temperature $T_{c}$  depends on the width $L$ of the square lattice. 
The continuum limit 
$\lim_{T \to \infty} m^{i} \longrightarrow 0$
is more closely approached with increasing lattice size $L$. Moreover, the phase transition appears to
occur more sharply as the system size $L$ increases for a given fixed parameter set $(G, g_{3})$. 
Again no discontinuous jump at the transition temperature $T_{c}$ is observed. 
We also examined the behavior of the energy histogram depicted in  Figure~\ref{fig3}. The energy probability distribution exhibits a smooth, continuous change with {\it no} double peak structure and all histograms maintain a similar shape throughout the temperature
interval $T \in [1.951, 2.118]$ with increments of $\Delta T = 0.001$ indicating that the transition is indeed of second order. 
\begin{figure}
    \includegraphics[width=1.0\columnwidth]{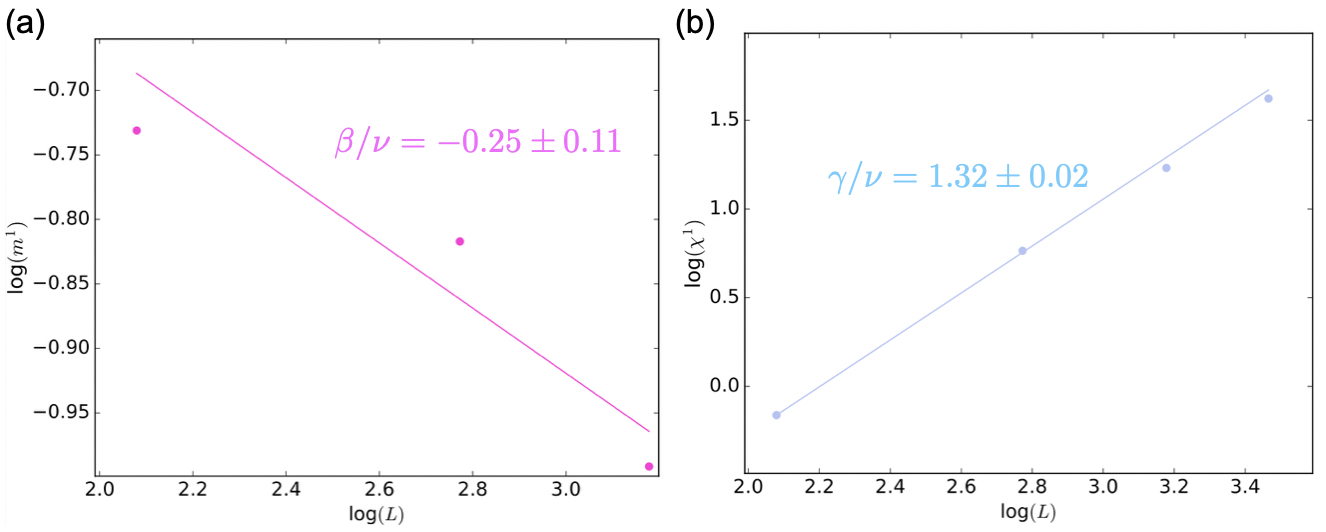}
\caption{ 
    Log-log plots of (a) the order parameter $m^{1}$ and (b) the susceptibility $\chi^{1}$ as a function of the lattice size $L$, at the critical point of the transition. The straight lines are the best fit to the data points. The respective slopes determine the critical exponents, i.e. $\gamma/\nu = 1.32 \,\pm\, 0.02$ and $\beta/\nu = 0.25 \,\pm\, 0.1$.  
}    
    \label{fig4ab}
\end{figure}

\subsection{MC results: Susceptibility and specific heat}\label{susheat}
We next examine the susceptibilities $\chi^{i}$, $i=1,2,3$, of the ordering CDW fields across temperature $T$, as shown in Fig.~\ref{fig3}(b). The susceptibility $\chi^{1}$ displays a characteristic trend, remaining near zero at both high and low temperatures $T$, with a pronounced peak emerging around the critical temperature $T_{c}$. The second susceptibility component, $\chi^{2}$, shown in Figures~\ref{app_fig2}(b) and~\ref{app_fig4}(b), follows a qualitatively similar pattern. As system size increases, the peak sharpens and grows, indicating a divergence at large sizes, particularly for $L = 32$.
   
Looking closer at the specific heat depicted in Figure~\ref{fig2}(c)  we notice a transition from $C = 1$ at low temperatures $T$ to a sharp peak in the critical temperature region and subsequently it decreases at higher temperatures. The height of the peak slightly varies with lattice size $L$. 

\begin{figure}
    \includegraphics[width=0.49\textwidth]{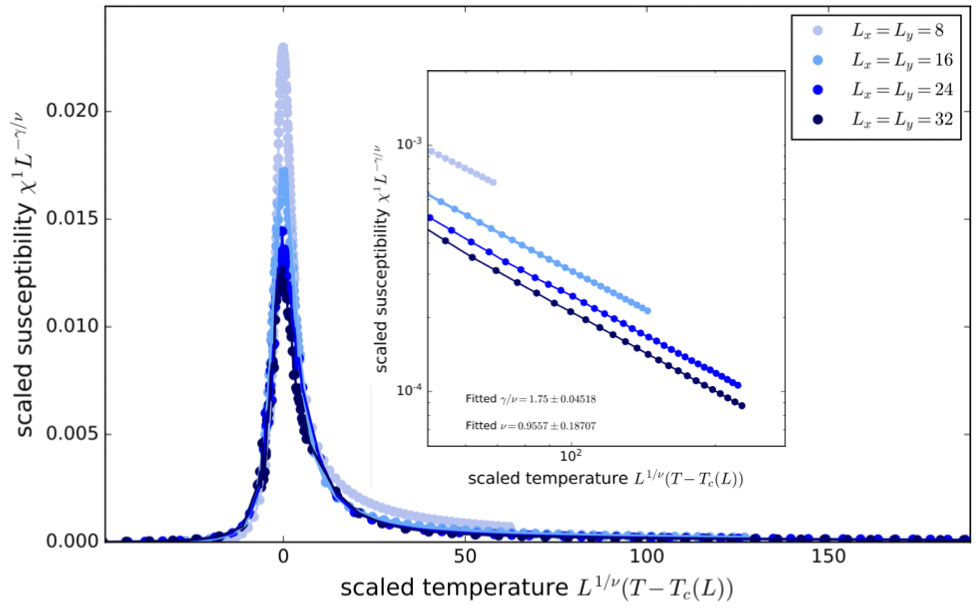}
    \caption{
     System parameters $(J,\Delta) = (1.0,1.0)$ and $(G,g_{3}) = (6.0,1.0)$. 
    (a) The scaling collapse for the susceptibility, i.e. $\chi^{1} L^{-\gamma/\nu}$ vs. $L^{1/\nu}(T-T_{c})$. 
    We see that the data for different system sizes $L_x=L_y= 8,16,24,32$ shows a reasonable collapse, except for the smallest system $L_x=L_y=8$. 
    (b) The same as in (a) but on a  log-log scale. The data points do not perfectly collapse. However, the discrepancy decreases for increasing system size. 
  }  
    \label{fig5}
\end{figure}

\section{Finite-size scaling}\label{finite}
We further analyzed the data obtained from the MC simulation by  performing finite-size scaling. Here we focus on the behavior of the magnetic susceptibility $\chi$ and the order parameter $m$ in the critical region of each lattice width $L$. We ran additional simulations with $\Delta T = 0.001$ in the critical region for each lattice size $L$ for three cases $(G,g_{3}) = (6.0,1.0)$, $(4.0,1.0)$ and $(2.0,1.0)$. 
The best finite-size scaling fit we found is consistent with the hypothesis of a second order (continuous) phase transition.  

To corroborate this, finite-size scaling relations are used 
to determine the values of the critical exponents. For instance, the finite-size scaling relations for the quantitites $m$ and $\chi$ as defined before in~\eqref{o_parameter} and~\eqref{mag_sus}, in the neighborhood of the critical point $T_{c}$, are 
\begin{align}
\begin{split}
\label{exponents}
     m^{i}_{L}(T_{c}) &=L^{-\beta/\nu} m^{i}_{0}(L^{1/\nu})
     \\
     \chi^{i}_{L}(T_{c}) &= L^{\gamma/\nu} \chi^{i}_{0}(L^{1/\nu})
\end{split}
\end{align}
with $i=1,2,3$, respectively. 
To make progress we determine the critical exponents of the model~\eqref{ham} from the Monte Carlo data. Subsequently, log-log plots of the Eqs.~\eqref{exponents} at the critical point $T_{c}$,  provide the critical exponents from the slope of the corresponding straight lines. This can be seen in Figure~\ref{fig4ab}, for the transition at $T_{c} \approx 1.951$. From the best fit to the data points we found $\beta/\nu = -0.25 \pm 0.11$ and $\gamma/\nu = 1.32 \pm 0.02$. 

To further improve the values found for the critical exponents we perform a scaling collapsing of the data points. We focus on the first component of the susceptibility $\chi^{1}$ and plot 
$\chi^{1} L^{-\gamma/\nu}$ over $L^{1/\nu}(T-T_{c})$ in Figure~\ref{fig5}. We observe that the data for different system sizes $L = 8, 16, 24, 32$ shows a reasonable collapse with parameters $\gamma = 1.75 \pm 0.045$ and $\nu = 0.956 \pm 0.187$ except for the smallest system of $L=8$. To clarify this, the inset provides a plot on log-log scale. While the data points for different system sizes $L$ do not perfectly collapse we note that the discrepancy decreases for increasing systems size. 

To close this section we mention that we performed additional MC simulations for two different parameter sets, namely $(G, g_3)= (4.0, 1.0)$ and $(G, g_3) = (2.0,1.0)$. The remaining parameters $J$ and $\Delta$ are again set to one, i.e. $J = \Delta =1.0$. 
We provide the obtained plots in the appendix. Note that the two additional parameter sets qualitatively do not differ from the parameter set $(G, g_3) = (6.0,1.0)$ discussed above. Indeed the results again indicate the the phase transition is of second order. 
\section{Conclusions}  
We have demonstrated two key findings regarding the CDWs in 2D kagome metals. First, the system undergoes a single phase transition from a disordered phase at high temperatures to an ordered phase where individual CDWs align. Second, this phase transition is significantly influenced by topological excitations, i.e. vortices, and exhibits characteristics of a second-order transition.
\section{Acknowledgements}
This work was supported by Office of Basic Energy Sciences, Material Sciences and Engineering Division, U.S. Department of Energy (DOE) under Contracts No. DE-SC0012704.
\appendix
\section{Additional Data for systems with parameters $(G, g_{3}) = (4.0, 1.0)$ and $(G, g_3) = (2.0,1.0)$ with $(J,\Delta) = (1.0,1.0)$}
\label{appA}
\begin{figure*}[b]
    \includegraphics[width=1.00\textwidth]{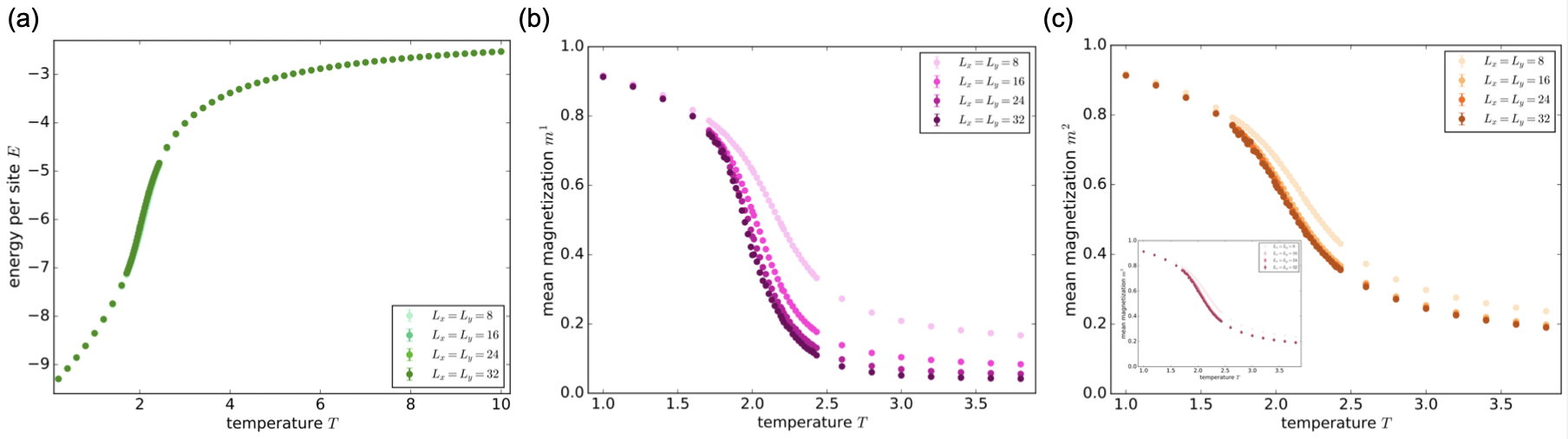}
\caption{
   (a) Energy per site $E$ as a function of temperature $T$ for the model in Eq.~\eqref{ham} with parameters $(G, g_3) = (2.0, 1.0)$ and $(J, \Delta) = (1.0, 1.0)$, shown for system sizes $L = 8, 16, 24, 32$. The continuous change in $E$ through the phase transition at $T_c$ is characteristic of a second-order transition.
   (b) Mean magnetization $m^1$ as a function of $T$, illustrating the continuous behavior of the order parameter near $T_c$. The value of $m^1$ above $T_c$ approaches zero as lattice size $L$ increases.
   (c) The second order parameter $m^2$ shows behavior similar to $m^1$, with $m^3$ (shown in the inset) following a comparable trend, collectively indicating a continuous second-order transition. 
}\label{app_fig1}
\end{figure*}

\begin{figure*}[b]
    \includegraphics[width=1.0\textwidth]{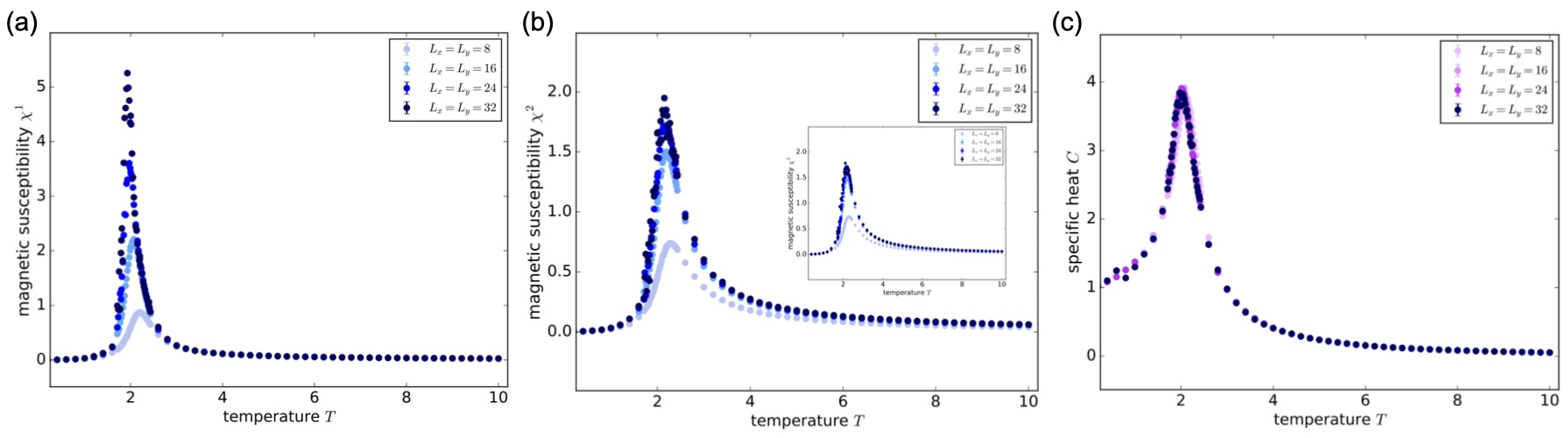}
\caption{
   (a) Magnetic susceptibility $\chi^1$ as a function of temperature $T$ shows a peak near the critical temperature $T_c$, indicating a second-order transition. The peak height increases with lattice size $L$. 
   (b) The second susceptibility component, $\chi^2$, and the third component (shown in the inset) exhibit similar behavior.
   (c) Specific heat $C$ as a function of $T$ rises from $C = 1$ at low temperatures, reaches a sharp peak near $T_c$, and decreases at higher temperatures, with the peak height varying slightly with lattice size.
}
    \label{app_fig2}
\end{figure*}

\begin{figure*}[b]
    \includegraphics[width=1.00\textwidth]{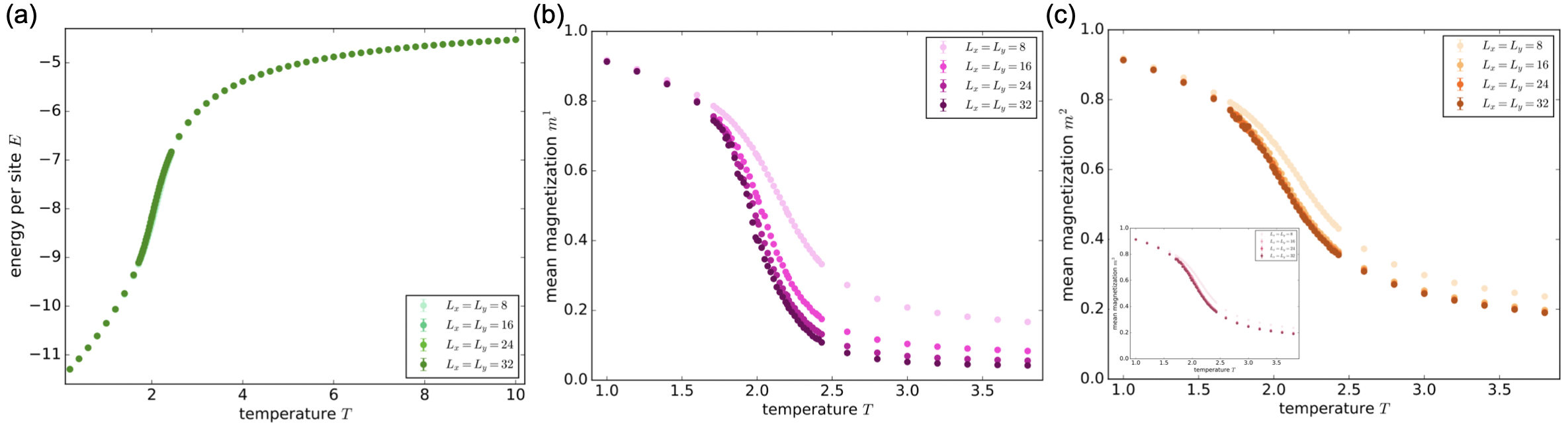}
\caption{
   (a) Energy per site $E$ versus temperature $T$ for the model in Eq.~\eqref{ham} with parameters $(G, g_3) = (4.0, 1.0)$ and $(J, \Delta) = (1.0, 1.0)$, shown for system sizes $L = 8, 16, 24, 32$. The continuous change in energy across the phase transition at $T_c$ indicates its second-order nature.
   (b) Mean magnetization $m^{1}$ as a function of $T$ displays the order parameter’s behavior near $T_c$. The continuous variation of $m^{1}$ signals a second-order transition, with $m^{1}$ approaching zero more closely for larger $L$ as $T$ increases beyond $T_c$.
   (c) The second order parameter $m^{2}$ shows similar behavior, with the third order parameter $m^{3}$ (shown in the inset) following this trend, collectively highlighting the continuous characteristics of a second-order phase transition.
}\label{app_fig3}
\end{figure*}

\begin{figure*}[b]
    \includegraphics[width=1.0\textwidth]{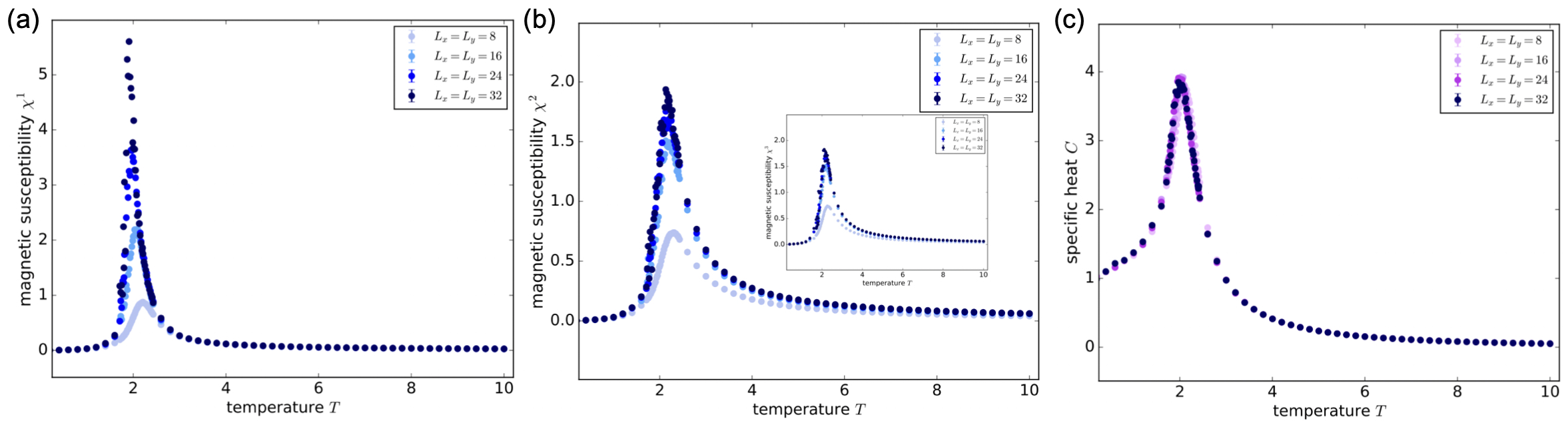}
\caption{
   (a) The magnetic susceptibility $\chi^{1}$ remains near zero at both high and low temperatures, but shows a pronounced peak around the critical temperature $T_c$, with the peak’s height increasing with lattice size $L$.
   (b) The inset displays the second susceptibility component, $\chi^{2}$, which shows a similar trend, as does the third component (also shown in the inset).
   (c) Specific heat $C$ versus temperature $T$ rises from $C = 1$ at low temperatures, reaching a sharp peak close to $T_c$ before decreasing again at higher temperatures. The peak’s height slightly depends on lattice size, highlighting the specific heat’s sensitivity to the phase transition. 
}
    \label{app_fig4}
\end{figure*}

In this Appendix we provide additional data for the cases that the ratio $G/g_3$ is smaller than it is for the system analyzed in the main text. Again we set $g_3 = 1.0$ and choose $G = 4.0$ and $G=2.0$, respectively. Results are displayed in Figures~\ref{app_fig1}-\ref{app_fig4}. 
To summarize, we note that  the energy per site $E$ plotted over the temperature $T$ exhibits the same continuous change in Figs.~\ref{app_fig1}(a) and ~\ref{app_fig3}(a) while its absolute value $|E|$, of course, increases with increasing coupling $G$. The order parameter $m^{i}$, $i=1,2,3$ displays exactly the same behavior for all three parameter sets presented in this work. Again its continuous nature across the temperature spectrum indicates a second order phase transition. 
Analyzing the susceptibility $\chi^{i}$, $i=1,2,3$ and the specific heat $C$ yields the same tendencies already discussed in Section~\ref{susheat} and ~\ref{finite}. 


\bibliographystyle{apsrev4-1}
\bibliography{main}

\begin{thebibliography}{39}%
\makeatletter
\providecommand \@ifxundefined [1]{%
 \@ifx{#1\undefined}
}%
\providecommand \@ifnum [1]{%
 \ifnum #1\expandafter \@firstoftwo
 \else \expandafter \@secondoftwo
 \fi
}%
\providecommand \@ifx [1]{%
 \ifx #1\expandafter \@firstoftwo
 \else \expandafter \@secondoftwo
 \fi
}%
\providecommand \natexlab [1]{#1}%
\providecommand \enquote  [1]{``#1''}%
\providecommand \bibnamefont  [1]{#1}%
\providecommand \bibfnamefont [1]{#1}%
\providecommand \citenamefont [1]{#1}%
\providecommand \href@noop [0]{\@secondoftwo}%
\providecommand \href [0]{\begingroup \@sanitize@url \@href}%
\providecommand \@href[1]{\@@startlink{#1}\@@href}%
\providecommand \@@href[1]{\endgroup#1\@@endlink}%
\providecommand \@sanitize@url [0]{\catcode `\\12\catcode `\$12\catcode `\&12\catcode `\#12\catcode `\^12\catcode `\_12\catcode `\%12\relax}%
\providecommand \@@startlink[1]{}%
\providecommand \@@endlink[0]{}%
\providecommand \url  [0]{\begingroup\@sanitize@url \@url }%
\providecommand \@url [1]{\endgroup\@href {#1}{\urlprefix }}%
\providecommand \urlprefix  [0]{URL }%
\providecommand \Eprint [0]{\href }%
\providecommand \doibase [0]{http://dx.doi.org/}%
\providecommand \selectlanguage [0]{\@gobble}%
\providecommand \bibinfo  [0]{\@secondoftwo}%
\providecommand \bibfield  [0]{\@secondoftwo}%
\providecommand \translation [1]{[#1]}%
\providecommand \BibitemOpen [0]{}%
\providecommand \bibitemStop [0]{}%
\providecommand \bibitemNoStop [0]{.\EOS\space}%
\providecommand \EOS [0]{\spacefactor3000\relax}%
\providecommand \BibitemShut  [1]{\csname bibitem#1\endcsname}%
\let\auto@bib@innerbib\@empty
\bibitem [{\citenamefont {Kennes}\ \emph {et~al.}(2021)\citenamefont {Kennes}, \citenamefont {Claassen}, \citenamefont {Xian}, \citenamefont {Georges}, \citenamefont {Millis}, \citenamefont {Hone}, \citenamefont {Dean}, \citenamefont {Basov}, \citenamefont {Pasupathy},\ and\ \citenamefont {Rubio}}]{kennes2021moire}%
  \BibitemOpen
  \bibfield  {author} {\bibinfo {author} {\bibfnamefont {D.~M.}\ \bibnamefont {Kennes}}, \bibinfo {author} {\bibfnamefont {M.}~\bibnamefont {Claassen}}, \bibinfo {author} {\bibfnamefont {L.}~\bibnamefont {Xian}}, \bibinfo {author} {\bibfnamefont {A.}~\bibnamefont {Georges}}, \bibinfo {author} {\bibfnamefont {A.~J.}\ \bibnamefont {Millis}}, \bibinfo {author} {\bibfnamefont {J.}~\bibnamefont {Hone}}, \bibinfo {author} {\bibfnamefont {C.~R.}\ \bibnamefont {Dean}}, \bibinfo {author} {\bibfnamefont {D.}~\bibnamefont {Basov}}, \bibinfo {author} {\bibfnamefont {A.~N.}\ \bibnamefont {Pasupathy}}, \ and\ \bibinfo {author} {\bibfnamefont {A.}~\bibnamefont {Rubio}},\ }\href@noop {} {\bibfield  {journal} {\bibinfo  {journal} {Nature Physics}\ }\textbf {\bibinfo {volume} {17}},\ \bibinfo {pages} {155} (\bibinfo {year} {2021})}\BibitemShut {NoStop}%
\bibitem [{\citenamefont {Dzero}\ \emph {et~al.}(2016)\citenamefont {Dzero}, \citenamefont {Xia}, \citenamefont {Galitski},\ and\ \citenamefont {Coleman}}]{dzero2016topological}%
  \BibitemOpen
  \bibfield  {author} {\bibinfo {author} {\bibfnamefont {M.}~\bibnamefont {Dzero}}, \bibinfo {author} {\bibfnamefont {J.}~\bibnamefont {Xia}}, \bibinfo {author} {\bibfnamefont {V.}~\bibnamefont {Galitski}}, \ and\ \bibinfo {author} {\bibfnamefont {P.}~\bibnamefont {Coleman}},\ }\href@noop {} {\bibfield  {journal} {\bibinfo  {journal} {Annual Review of Condensed Matter Physics}\ }\textbf {\bibinfo {volume} {7}},\ \bibinfo {pages} {249} (\bibinfo {year} {2016})}\BibitemShut {NoStop}%
\bibitem [{\citenamefont {Ortiz}\ \emph {et~al.}(2019{\natexlab{a}})\citenamefont {Ortiz}, \citenamefont {Gomes}, \citenamefont {Morey}, \citenamefont {Winiarski}, \citenamefont {Bordelon}, \citenamefont {Mangum}, \citenamefont {Oswald}, \citenamefont {Rodriguez-Rivera}, \citenamefont {Neilson}, \citenamefont {Wilson}, \citenamefont {Ertekin}, \citenamefont {McQueen},\ and\ \citenamefont {Toberer}}]{Gomes}%
  \BibitemOpen
  \bibfield  {author} {\bibinfo {author} {\bibfnamefont {B.~R.}\ \bibnamefont {Ortiz}}, \bibinfo {author} {\bibfnamefont {L.~C.}\ \bibnamefont {Gomes}}, \bibinfo {author} {\bibfnamefont {J.~R.}\ \bibnamefont {Morey}}, \bibinfo {author} {\bibfnamefont {M.}~\bibnamefont {Winiarski}}, \bibinfo {author} {\bibfnamefont {M.}~\bibnamefont {Bordelon}}, \bibinfo {author} {\bibfnamefont {J.~S.}\ \bibnamefont {Mangum}}, \bibinfo {author} {\bibfnamefont {I.~W.~H.}\ \bibnamefont {Oswald}}, \bibinfo {author} {\bibfnamefont {J.~A.}\ \bibnamefont {Rodriguez-Rivera}}, \bibinfo {author} {\bibfnamefont {J.~R.}\ \bibnamefont {Neilson}}, \bibinfo {author} {\bibfnamefont {S.~D.}\ \bibnamefont {Wilson}}, \bibinfo {author} {\bibfnamefont {E.}~\bibnamefont {Ertekin}}, \bibinfo {author} {\bibfnamefont {T.~M.}\ \bibnamefont {McQueen}}, \ and\ \bibinfo {author} {\bibfnamefont {E.~S.}\ \bibnamefont {Toberer}},\ }\href {\doibase 10.1103/PhysRevMaterials.3.094407} {\bibfield  {journal} {\bibinfo  {journal} {Phys. Rev. Mater.}\ }\textbf
  {\bibinfo {volume} {3}},\ \bibinfo {pages} {094407} (\bibinfo {year} {2019}{\natexlab{a}})}\BibitemShut {NoStop}%
\bibitem [{\citenamefont {Ortiz}\ \emph {et~al.}(2020)\citenamefont {Ortiz}, \citenamefont {Teicher}, \citenamefont {Hu}, \citenamefont {Zuo}, \citenamefont {Sarte}, \citenamefont {Schueller}, \citenamefont {Abeykoon}, \citenamefont {Krogstad}, \citenamefont {Rosenkranz}, \citenamefont {Osborn}, \citenamefont {Seshadri}, \citenamefont {Balents}, \citenamefont {He},\ and\ \citenamefont {Wilson}}]{Teicher}%
  \BibitemOpen
  \bibfield  {author} {\bibinfo {author} {\bibfnamefont {B.~R.}\ \bibnamefont {Ortiz}}, \bibinfo {author} {\bibfnamefont {S.~M.~L.}\ \bibnamefont {Teicher}}, \bibinfo {author} {\bibfnamefont {Y.}~\bibnamefont {Hu}}, \bibinfo {author} {\bibfnamefont {J.~L.}\ \bibnamefont {Zuo}}, \bibinfo {author} {\bibfnamefont {P.~M.}\ \bibnamefont {Sarte}}, \bibinfo {author} {\bibfnamefont {E.~C.}\ \bibnamefont {Schueller}}, \bibinfo {author} {\bibfnamefont {A.~M.~M.}\ \bibnamefont {Abeykoon}}, \bibinfo {author} {\bibfnamefont {M.~J.}\ \bibnamefont {Krogstad}}, \bibinfo {author} {\bibfnamefont {S.}~\bibnamefont {Rosenkranz}}, \bibinfo {author} {\bibfnamefont {R.}~\bibnamefont {Osborn}}, \bibinfo {author} {\bibfnamefont {R.}~\bibnamefont {Seshadri}}, \bibinfo {author} {\bibfnamefont {L.}~\bibnamefont {Balents}}, \bibinfo {author} {\bibfnamefont {J.}~\bibnamefont {He}}, \ and\ \bibinfo {author} {\bibfnamefont {S.~D.}\ \bibnamefont {Wilson}},\ }\href {\doibase 10.1103/PhysRevLett.125.247002} {\bibfield  {journal} {\bibinfo
  {journal} {Phys. Rev. Lett.}\ }\textbf {\bibinfo {volume} {125}},\ \bibinfo {pages} {247002} (\bibinfo {year} {2020})}\BibitemShut {NoStop}%
\bibitem [{\citenamefont {Ortiz}\ \emph {et~al.}(2021{\natexlab{a}})\citenamefont {Ortiz}, \citenamefont {Sarte}, \citenamefont {Kenney}, \citenamefont {Graf}, \citenamefont {Teicher}, \citenamefont {Seshadri},\ and\ \citenamefont {Wilson}}]{Sarte}%
  \BibitemOpen
  \bibfield  {author} {\bibinfo {author} {\bibfnamefont {B.~R.}\ \bibnamefont {Ortiz}}, \bibinfo {author} {\bibfnamefont {P.~M.}\ \bibnamefont {Sarte}}, \bibinfo {author} {\bibfnamefont {E.~M.}\ \bibnamefont {Kenney}}, \bibinfo {author} {\bibfnamefont {M.~J.}\ \bibnamefont {Graf}}, \bibinfo {author} {\bibfnamefont {S.~M.~L.}\ \bibnamefont {Teicher}}, \bibinfo {author} {\bibfnamefont {R.}~\bibnamefont {Seshadri}}, \ and\ \bibinfo {author} {\bibfnamefont {S.~D.}\ \bibnamefont {Wilson}},\ }\href {\doibase 10.1103/PhysRevMaterials.5.034801} {\bibfield  {journal} {\bibinfo  {journal} {Phys. Rev. Mater.}\ }\textbf {\bibinfo {volume} {5}},\ \bibinfo {pages} {034801} (\bibinfo {year} {2021}{\natexlab{a}})}\BibitemShut {NoStop}%
\bibitem [{\citenamefont {Jiang}\ \emph {et~al.}(2021)\citenamefont {Jiang}, \citenamefont {Yin}, \citenamefont {Denner}, \citenamefont {Shumiya}, \citenamefont {Ortiz}, \citenamefont {Xu}, \citenamefont {Guguchia}, \citenamefont {He}, \citenamefont {Hossain}, \citenamefont {Liu} \emph {et~al.}}]{jiang2021unconventional}%
  \BibitemOpen
  \bibfield  {author} {\bibinfo {author} {\bibfnamefont {Y.-X.}\ \bibnamefont {Jiang}}, \bibinfo {author} {\bibfnamefont {J.-X.}\ \bibnamefont {Yin}}, \bibinfo {author} {\bibfnamefont {M.~M.}\ \bibnamefont {Denner}}, \bibinfo {author} {\bibfnamefont {N.}~\bibnamefont {Shumiya}}, \bibinfo {author} {\bibfnamefont {B.~R.}\ \bibnamefont {Ortiz}}, \bibinfo {author} {\bibfnamefont {G.}~\bibnamefont {Xu}}, \bibinfo {author} {\bibfnamefont {Z.}~\bibnamefont {Guguchia}}, \bibinfo {author} {\bibfnamefont {J.}~\bibnamefont {He}}, \bibinfo {author} {\bibfnamefont {M.~S.}\ \bibnamefont {Hossain}}, \bibinfo {author} {\bibfnamefont {X.}~\bibnamefont {Liu}},  \emph {et~al.},\ }\href@noop {} {\bibfield  {journal} {\bibinfo  {journal} {Nature materials}\ }\textbf {\bibinfo {volume} {20}},\ \bibinfo {pages} {1353} (\bibinfo {year} {2021})}\BibitemShut {NoStop}%
\bibitem [{\citenamefont {Li}\ \emph {et~al.}(2021)\citenamefont {Li}, \citenamefont {Zhang}, \citenamefont {Yilmaz}, \citenamefont {Pai}, \citenamefont {Marvinney}, \citenamefont {Said}, \citenamefont {Yin}, \citenamefont {Gong}, \citenamefont {Tu}, \citenamefont {Vescovo}, \citenamefont {Nelson}, \citenamefont {Moore}, \citenamefont {Murakami}, \citenamefont {Lei}, \citenamefont {Lee}, \citenamefont {Lawrie},\ and\ \citenamefont {Miao}}]{Li}%
  \BibitemOpen
  \bibfield  {author} {\bibinfo {author} {\bibfnamefont {H.}~\bibnamefont {Li}}, \bibinfo {author} {\bibfnamefont {T.~T.}\ \bibnamefont {Zhang}}, \bibinfo {author} {\bibfnamefont {T.}~\bibnamefont {Yilmaz}}, \bibinfo {author} {\bibfnamefont {Y.~Y.}\ \bibnamefont {Pai}}, \bibinfo {author} {\bibfnamefont {C.~E.}\ \bibnamefont {Marvinney}}, \bibinfo {author} {\bibfnamefont {A.}~\bibnamefont {Said}}, \bibinfo {author} {\bibfnamefont {Q.~W.}\ \bibnamefont {Yin}}, \bibinfo {author} {\bibfnamefont {C.~S.}\ \bibnamefont {Gong}}, \bibinfo {author} {\bibfnamefont {Z.~J.}\ \bibnamefont {Tu}}, \bibinfo {author} {\bibfnamefont {E.}~\bibnamefont {Vescovo}}, \bibinfo {author} {\bibfnamefont {C.~S.}\ \bibnamefont {Nelson}}, \bibinfo {author} {\bibfnamefont {R.~G.}\ \bibnamefont {Moore}}, \bibinfo {author} {\bibfnamefont {S.}~\bibnamefont {Murakami}}, \bibinfo {author} {\bibfnamefont {H.~C.}\ \bibnamefont {Lei}}, \bibinfo {author} {\bibfnamefont {H.~N.}\ \bibnamefont {Lee}}, \bibinfo {author} {\bibfnamefont {B.~J.}\
  \bibnamefont {Lawrie}}, \ and\ \bibinfo {author} {\bibfnamefont {H.}~\bibnamefont {Miao}},\ }\href {\doibase 10.1103/PhysRevX.11.031050} {\bibfield  {journal} {\bibinfo  {journal} {Phys. Rev. X}\ }\textbf {\bibinfo {volume} {11}},\ \bibinfo {pages} {031050} (\bibinfo {year} {2021})}\BibitemShut {NoStop}%
\bibitem [{\citenamefont {Uykur}\ \emph {et~al.}(2022)\citenamefont {Uykur}, \citenamefont {Ortiz}, \citenamefont {Wilson}, \citenamefont {Dressel},\ and\ \citenamefont {Tsirlin}}]{uykur2022optical}%
  \BibitemOpen
  \bibfield  {author} {\bibinfo {author} {\bibfnamefont {E.}~\bibnamefont {Uykur}}, \bibinfo {author} {\bibfnamefont {B.~R.}\ \bibnamefont {Ortiz}}, \bibinfo {author} {\bibfnamefont {S.~D.}\ \bibnamefont {Wilson}}, \bibinfo {author} {\bibfnamefont {M.}~\bibnamefont {Dressel}}, \ and\ \bibinfo {author} {\bibfnamefont {A.~A.}\ \bibnamefont {Tsirlin}},\ }\href@noop {} {\bibfield  {journal} {\bibinfo  {journal} {npj Quantum Materials}\ }\textbf {\bibinfo {volume} {7}},\ \bibinfo {pages} {16} (\bibinfo {year} {2022})}\BibitemShut {NoStop}%
\bibitem [{\citenamefont {Ortiz}\ \emph {et~al.}(2021{\natexlab{b}})\citenamefont {Ortiz}, \citenamefont {Teicher}, \citenamefont {Kautzsch}, \citenamefont {Sarte}, \citenamefont {Ratcliff}, \citenamefont {Harter}, \citenamefont {Ruff}, \citenamefont {Seshadri},\ and\ \citenamefont {Wilson}}]{Ruff}%
  \BibitemOpen
  \bibfield  {author} {\bibinfo {author} {\bibfnamefont {B.~R.}\ \bibnamefont {Ortiz}}, \bibinfo {author} {\bibfnamefont {S.~M.~L.}\ \bibnamefont {Teicher}}, \bibinfo {author} {\bibfnamefont {L.}~\bibnamefont {Kautzsch}}, \bibinfo {author} {\bibfnamefont {P.~M.}\ \bibnamefont {Sarte}}, \bibinfo {author} {\bibfnamefont {N.}~\bibnamefont {Ratcliff}}, \bibinfo {author} {\bibfnamefont {J.}~\bibnamefont {Harter}}, \bibinfo {author} {\bibfnamefont {J.~P.~C.}\ \bibnamefont {Ruff}}, \bibinfo {author} {\bibfnamefont {R.}~\bibnamefont {Seshadri}}, \ and\ \bibinfo {author} {\bibfnamefont {S.~D.}\ \bibnamefont {Wilson}},\ }\href {\doibase 10.1103/PhysRevX.11.041030} {\bibfield  {journal} {\bibinfo  {journal} {Phys. Rev. X}\ }\textbf {\bibinfo {volume} {11}},\ \bibinfo {pages} {041030} (\bibinfo {year} {2021}{\natexlab{b}})}\BibitemShut {NoStop}%
\bibitem [{\citenamefont {Tan}\ \emph {et~al.}(2021)\citenamefont {Tan}, \citenamefont {Liu}, \citenamefont {Wang},\ and\ \citenamefont {Yan}}]{Tan}%
  \BibitemOpen
  \bibfield  {author} {\bibinfo {author} {\bibfnamefont {H.}~\bibnamefont {Tan}}, \bibinfo {author} {\bibfnamefont {Y.}~\bibnamefont {Liu}}, \bibinfo {author} {\bibfnamefont {Z.}~\bibnamefont {Wang}}, \ and\ \bibinfo {author} {\bibfnamefont {B.}~\bibnamefont {Yan}},\ }\href {\doibase 10.1103/PhysRevLett.127.046401} {\bibfield  {journal} {\bibinfo  {journal} {Phys. Rev. Lett.}\ }\textbf {\bibinfo {volume} {127}},\ \bibinfo {pages} {046401} (\bibinfo {year} {2021})}\BibitemShut {NoStop}%
\bibitem [{\citenamefont {Denner}\ \emph {et~al.}(2021{\natexlab{a}})\citenamefont {Denner}, \citenamefont {Thomale},\ and\ \citenamefont {Neupert}}]{Denner}%
  \BibitemOpen
  \bibfield  {author} {\bibinfo {author} {\bibfnamefont {M.~M.}\ \bibnamefont {Denner}}, \bibinfo {author} {\bibfnamefont {R.}~\bibnamefont {Thomale}}, \ and\ \bibinfo {author} {\bibfnamefont {T.}~\bibnamefont {Neupert}},\ }\href {\doibase 10.1103/PhysRevLett.127.217601} {\bibfield  {journal} {\bibinfo  {journal} {Phys. Rev. Lett.}\ }\textbf {\bibinfo {volume} {127}},\ \bibinfo {pages} {217601} (\bibinfo {year} {2021}{\natexlab{a}})}\BibitemShut {NoStop}%
\bibitem [{\citenamefont {Feng}\ \emph {et~al.}(2021)\citenamefont {Feng}, \citenamefont {Jiang}, \citenamefont {Wang},\ and\ \citenamefont {Hu}}]{feng2021chiral}%
  \BibitemOpen
  \bibfield  {author} {\bibinfo {author} {\bibfnamefont {X.}~\bibnamefont {Feng}}, \bibinfo {author} {\bibfnamefont {K.}~\bibnamefont {Jiang}}, \bibinfo {author} {\bibfnamefont {Z.}~\bibnamefont {Wang}}, \ and\ \bibinfo {author} {\bibfnamefont {J.}~\bibnamefont {Hu}},\ }\href@noop {} {\bibfield  {journal} {\bibinfo  {journal} {Science bulletin}\ }\textbf {\bibinfo {volume} {66}},\ \bibinfo {pages} {1384} (\bibinfo {year} {2021})}\BibitemShut {NoStop}%
\bibitem [{\citenamefont {Yu}\ \emph {et~al.}()\citenamefont {Yu}, \citenamefont {Wang}, \citenamefont {Zhang}, \citenamefont {Sander}, \citenamefont {Ni}, \citenamefont {Lu}, \citenamefont {Ma}, \citenamefont {Wang}, \citenamefont {Zhao}, \citenamefont {Chen} \emph {et~al.}}]{yu2107evidence}%
  \BibitemOpen
  \bibfield  {author} {\bibinfo {author} {\bibfnamefont {L.}~\bibnamefont {Yu}}, \bibinfo {author} {\bibfnamefont {C.}~\bibnamefont {Wang}}, \bibinfo {author} {\bibfnamefont {Y.}~\bibnamefont {Zhang}}, \bibinfo {author} {\bibfnamefont {M.}~\bibnamefont {Sander}}, \bibinfo {author} {\bibfnamefont {S.}~\bibnamefont {Ni}}, \bibinfo {author} {\bibfnamefont {Z.}~\bibnamefont {Lu}}, \bibinfo {author} {\bibfnamefont {S.}~\bibnamefont {Ma}}, \bibinfo {author} {\bibfnamefont {Z.}~\bibnamefont {Wang}}, \bibinfo {author} {\bibfnamefont {Z.}~\bibnamefont {Zhao}}, \bibinfo {author} {\bibfnamefont {H.}~\bibnamefont {Chen}},  \emph {et~al.},\ }\href@noop {} {\bibinfo  {journal} {arXiv preprint arXiv:2107.10714}\ }\BibitemShut {NoStop}%
\bibitem [{\citenamefont {Yang}\ \emph {et~al.}(2020)\citenamefont {Yang}, \citenamefont {Wang}, \citenamefont {Ortiz}, \citenamefont {Liu}, \citenamefont {Gayles}, \citenamefont {Derunova}, \citenamefont {Gonzalez-Hernandez}, \citenamefont {{\v{S}}mejkal}, \citenamefont {Chen}, \citenamefont {Parkin} \emph {et~al.}}]{yang2020giant}%
  \BibitemOpen
\bibfield  {journal} {  }\bibfield  {author} {\bibinfo {author} {\bibfnamefont {S.-Y.}\ \bibnamefont {Yang}}, \bibinfo {author} {\bibfnamefont {Y.}~\bibnamefont {Wang}}, \bibinfo {author} {\bibfnamefont {B.~R.}\ \bibnamefont {Ortiz}}, \bibinfo {author} {\bibfnamefont {D.}~\bibnamefont {Liu}}, \bibinfo {author} {\bibfnamefont {J.}~\bibnamefont {Gayles}}, \bibinfo {author} {\bibfnamefont {E.}~\bibnamefont {Derunova}}, \bibinfo {author} {\bibfnamefont {R.}~\bibnamefont {Gonzalez-Hernandez}}, \bibinfo {author} {\bibfnamefont {L.}~\bibnamefont {{\v{S}}mejkal}}, \bibinfo {author} {\bibfnamefont {Y.}~\bibnamefont {Chen}}, \bibinfo {author} {\bibfnamefont {S.~S.}\ \bibnamefont {Parkin}},  \emph {et~al.},\ }\href@noop {} {\bibfield  {journal} {\bibinfo  {journal} {Science advances}\ }\textbf {\bibinfo {volume} {6}},\ \bibinfo {pages} {eabb6003} (\bibinfo {year} {2020})}\BibitemShut {NoStop}%
\bibitem [{\citenamefont {Yu}\ \emph {et~al.}(2021)\citenamefont {Yu}, \citenamefont {Wu}, \citenamefont {Wang}, \citenamefont {Lei}, \citenamefont {Zhuo}, \citenamefont {Ying},\ and\ \citenamefont {Chen}}]{YuWu}%
  \BibitemOpen
  \bibfield  {author} {\bibinfo {author} {\bibfnamefont {F.~H.}\ \bibnamefont {Yu}}, \bibinfo {author} {\bibfnamefont {T.}~\bibnamefont {Wu}}, \bibinfo {author} {\bibfnamefont {Z.~Y.}\ \bibnamefont {Wang}}, \bibinfo {author} {\bibfnamefont {B.}~\bibnamefont {Lei}}, \bibinfo {author} {\bibfnamefont {W.~Z.}\ \bibnamefont {Zhuo}}, \bibinfo {author} {\bibfnamefont {J.~J.}\ \bibnamefont {Ying}}, \ and\ \bibinfo {author} {\bibfnamefont {X.~H.}\ \bibnamefont {Chen}},\ }\href {\doibase 10.1103/PhysRevB.104.L041103} {\bibfield  {journal} {\bibinfo  {journal} {Phys. Rev. B}\ }\textbf {\bibinfo {volume} {104}},\ \bibinfo {pages} {L041103} (\bibinfo {year} {2021})}\BibitemShut {NoStop}%
\bibitem [{\citenamefont {Yin}\ \emph {et~al.}(2022)\citenamefont {Yin}, \citenamefont {Lian},\ and\ \citenamefont {Hasan}}]{yin2022topological}%
  \BibitemOpen
  \bibfield  {author} {\bibinfo {author} {\bibfnamefont {J.-X.}\ \bibnamefont {Yin}}, \bibinfo {author} {\bibfnamefont {B.}~\bibnamefont {Lian}}, \ and\ \bibinfo {author} {\bibfnamefont {M.~Z.}\ \bibnamefont {Hasan}},\ }\href@noop {} {\bibfield  {journal} {\bibinfo  {journal} {Nature}\ }\textbf {\bibinfo {volume} {612}},\ \bibinfo {pages} {647} (\bibinfo {year} {2022})}\BibitemShut {NoStop}%
\bibitem [{\citenamefont {Ortiz}\ \emph {et~al.}(2019{\natexlab{b}})\citenamefont {Ortiz}, \citenamefont {Gomes}, \citenamefont {Morey}, \citenamefont {Winiarski}, \citenamefont {Bordelon}, \citenamefont {Mangum}, \citenamefont {Oswald}, \citenamefont {Rodriguez-Rivera}, \citenamefont {Neilson}, \citenamefont {Wilson}, \citenamefont {Ertekin}, \citenamefont {McQueen},\ and\ \citenamefont {Toberer}}]{OrtizB2019}%
  \BibitemOpen
  \bibfield  {author} {\bibinfo {author} {\bibfnamefont {B.~R.}\ \bibnamefont {Ortiz}}, \bibinfo {author} {\bibfnamefont {L.~C.}\ \bibnamefont {Gomes}}, \bibinfo {author} {\bibfnamefont {J.~R.}\ \bibnamefont {Morey}}, \bibinfo {author} {\bibfnamefont {M.}~\bibnamefont {Winiarski}}, \bibinfo {author} {\bibfnamefont {M.}~\bibnamefont {Bordelon}}, \bibinfo {author} {\bibfnamefont {J.~S.}\ \bibnamefont {Mangum}}, \bibinfo {author} {\bibfnamefont {I.~W.~H.}\ \bibnamefont {Oswald}}, \bibinfo {author} {\bibfnamefont {J.~A.}\ \bibnamefont {Rodriguez-Rivera}}, \bibinfo {author} {\bibfnamefont {J.~R.}\ \bibnamefont {Neilson}}, \bibinfo {author} {\bibfnamefont {S.~D.}\ \bibnamefont {Wilson}}, \bibinfo {author} {\bibfnamefont {E.}~\bibnamefont {Ertekin}}, \bibinfo {author} {\bibfnamefont {T.~M.}\ \bibnamefont {McQueen}}, \ and\ \bibinfo {author} {\bibfnamefont {E.~S.}\ \bibnamefont {Toberer}},\ }\href {\doibase 10.1103/PhysRevMaterials.3.094407} {\bibfield  {journal} {\bibinfo  {journal} {Phys. Rev. Mater.}\ }\textbf
  {\bibinfo {volume} {3}},\ \bibinfo {pages} {094407} (\bibinfo {year} {2019}{\natexlab{b}})}\BibitemShut {NoStop}%
\bibitem [{\citenamefont {Ortiz}\ \emph {et~al.}(2019{\natexlab{c}})\citenamefont {Ortiz}, \citenamefont {Gomes}, \citenamefont {Morey}, \citenamefont {Winiarski}, \citenamefont {Bordelon}, \citenamefont {Mangum}, \citenamefont {Oswald}, \citenamefont {Rodriguez-Rivera}, \citenamefont {Neilson}, \citenamefont {Wilson} \emph {et~al.}}]{ortiz2019new}%
  \BibitemOpen
  \bibfield  {author} {\bibinfo {author} {\bibfnamefont {B.~R.}\ \bibnamefont {Ortiz}}, \bibinfo {author} {\bibfnamefont {L.~C.}\ \bibnamefont {Gomes}}, \bibinfo {author} {\bibfnamefont {J.~R.}\ \bibnamefont {Morey}}, \bibinfo {author} {\bibfnamefont {M.}~\bibnamefont {Winiarski}}, \bibinfo {author} {\bibfnamefont {M.}~\bibnamefont {Bordelon}}, \bibinfo {author} {\bibfnamefont {J.~S.}\ \bibnamefont {Mangum}}, \bibinfo {author} {\bibfnamefont {I.~W.}\ \bibnamefont {Oswald}}, \bibinfo {author} {\bibfnamefont {J.~A.}\ \bibnamefont {Rodriguez-Rivera}}, \bibinfo {author} {\bibfnamefont {J.~R.}\ \bibnamefont {Neilson}}, \bibinfo {author} {\bibfnamefont {S.~D.}\ \bibnamefont {Wilson}},  \emph {et~al.},\ }\href@noop {} {\bibfield  {journal} {\bibinfo  {journal} {Physical Review Materials}\ }\textbf {\bibinfo {volume} {3}},\ \bibinfo {pages} {094407} (\bibinfo {year} {2019}{\natexlab{c}})}\BibitemShut {NoStop}%
\bibitem [{\citenamefont {Chen}\ \emph {et~al.}(2021)\citenamefont {Chen}, \citenamefont {Yang}, \citenamefont {Hu}, \citenamefont {Zhao}, \citenamefont {Yuan}, \citenamefont {Xing}, \citenamefont {Qian}, \citenamefont {Huang}, \citenamefont {Li}, \citenamefont {Ye} \emph {et~al.}}]{chen2021roton}%
  \BibitemOpen
  \bibfield  {author} {\bibinfo {author} {\bibfnamefont {H.}~\bibnamefont {Chen}}, \bibinfo {author} {\bibfnamefont {H.}~\bibnamefont {Yang}}, \bibinfo {author} {\bibfnamefont {B.}~\bibnamefont {Hu}}, \bibinfo {author} {\bibfnamefont {Z.}~\bibnamefont {Zhao}}, \bibinfo {author} {\bibfnamefont {J.}~\bibnamefont {Yuan}}, \bibinfo {author} {\bibfnamefont {Y.}~\bibnamefont {Xing}}, \bibinfo {author} {\bibfnamefont {G.}~\bibnamefont {Qian}}, \bibinfo {author} {\bibfnamefont {Z.}~\bibnamefont {Huang}}, \bibinfo {author} {\bibfnamefont {G.}~\bibnamefont {Li}}, \bibinfo {author} {\bibfnamefont {Y.}~\bibnamefont {Ye}},  \emph {et~al.},\ }\href@noop {} {\bibfield  {journal} {\bibinfo  {journal} {Nature}\ }\textbf {\bibinfo {volume} {599}},\ \bibinfo {pages} {222} (\bibinfo {year} {2021})}\BibitemShut {NoStop}%
\bibitem [{\citenamefont {Li}\ \emph {et~al.}(2023)\citenamefont {Li}, \citenamefont {Oh}, \citenamefont {Kang}, \citenamefont {Zhao}, \citenamefont {Ortiz}, \citenamefont {Oey}, \citenamefont {Fang}, \citenamefont {Ren}, \citenamefont {Jozwiak}, \citenamefont {Bostwick}, \citenamefont {Rotenberg}, \citenamefont {Checkelsky}, \citenamefont {Wang}, \citenamefont {Wilson}, \citenamefont {Comin},\ and\ \citenamefont {Zeljkovic}}]{PDWLiPRX2023}%
  \BibitemOpen
  \bibfield  {author} {\bibinfo {author} {\bibfnamefont {H.}~\bibnamefont {Li}}, \bibinfo {author} {\bibfnamefont {D.}~\bibnamefont {Oh}}, \bibinfo {author} {\bibfnamefont {M.}~\bibnamefont {Kang}}, \bibinfo {author} {\bibfnamefont {H.}~\bibnamefont {Zhao}}, \bibinfo {author} {\bibfnamefont {B.~R.}\ \bibnamefont {Ortiz}}, \bibinfo {author} {\bibfnamefont {Y.}~\bibnamefont {Oey}}, \bibinfo {author} {\bibfnamefont {S.}~\bibnamefont {Fang}}, \bibinfo {author} {\bibfnamefont {Z.}~\bibnamefont {Ren}}, \bibinfo {author} {\bibfnamefont {C.}~\bibnamefont {Jozwiak}}, \bibinfo {author} {\bibfnamefont {A.}~\bibnamefont {Bostwick}}, \bibinfo {author} {\bibfnamefont {E.}~\bibnamefont {Rotenberg}}, \bibinfo {author} {\bibfnamefont {J.~G.}\ \bibnamefont {Checkelsky}}, \bibinfo {author} {\bibfnamefont {Z.}~\bibnamefont {Wang}}, \bibinfo {author} {\bibfnamefont {S.~D.}\ \bibnamefont {Wilson}}, \bibinfo {author} {\bibfnamefont {R.}~\bibnamefont {Comin}}, \ and\ \bibinfo {author} {\bibfnamefont {I.}~\bibnamefont {Zeljkovic}},\
  }\href {\doibase 10.1103/PhysRevX.13.031030} {\bibfield  {journal} {\bibinfo  {journal} {Phys. Rev. X}\ }\textbf {\bibinfo {volume} {13}},\ \bibinfo {pages} {031030} (\bibinfo {year} {2023})}\BibitemShut {NoStop}%
\bibitem [{\citenamefont {Tsvelik}\ and\ \citenamefont {Sarkar}(2023)}]{Sarkar2023}%
  \BibitemOpen
  \bibfield  {author} {\bibinfo {author} {\bibfnamefont {A.~M.}\ \bibnamefont {Tsvelik}}\ and\ \bibinfo {author} {\bibfnamefont {S.}~\bibnamefont {Sarkar}},\ }\href {\doibase 10.1103/PhysRevB.108.045119} {\bibfield  {journal} {\bibinfo  {journal} {Phys.Rev. B}\ }\textbf {\bibinfo {volume} {108}},\ \bibinfo {pages} {045119} (\bibinfo {year} {2023})}\BibitemShut {NoStop}%
\bibitem [{\citenamefont {Denner}\ \emph {et~al.}(2021{\natexlab{b}})\citenamefont {Denner}, \citenamefont {Thomale},\ and\ \citenamefont {Neupert}}]{TRS_ThomalePRL2021}%
  \BibitemOpen
  \bibfield  {author} {\bibinfo {author} {\bibfnamefont {M.~M.}\ \bibnamefont {Denner}}, \bibinfo {author} {\bibfnamefont {R.}~\bibnamefont {Thomale}}, \ and\ \bibinfo {author} {\bibfnamefont {T.}~\bibnamefont {Neupert}},\ }\href {\doibase 10.1103/PhysRevLett.127.217601} {\bibfield  {journal} {\bibinfo  {journal} {Phys. Rev. Lett.}\ }\textbf {\bibinfo {volume} {127}},\ \bibinfo {pages} {217601} (\bibinfo {year} {2021}{\natexlab{b}})}\BibitemShut {NoStop}%
\bibitem [{\citenamefont {Mielke~III}\ \emph {et~al.}(2022)\citenamefont {Mielke~III}, \citenamefont {Das}, \citenamefont {Yin}, \citenamefont {Liu}, \citenamefont {Gupta}, \citenamefont {Jiang}, \citenamefont {Medarde}, \citenamefont {Wu}, \citenamefont {Lei}, \citenamefont {Chang} \emph {et~al.}}]{mielke2022time}%
  \BibitemOpen
  \bibfield  {author} {\bibinfo {author} {\bibfnamefont {C.}~\bibnamefont {Mielke~III}}, \bibinfo {author} {\bibfnamefont {D.}~\bibnamefont {Das}}, \bibinfo {author} {\bibfnamefont {J.-X.}\ \bibnamefont {Yin}}, \bibinfo {author} {\bibfnamefont {H.}~\bibnamefont {Liu}}, \bibinfo {author} {\bibfnamefont {R.}~\bibnamefont {Gupta}}, \bibinfo {author} {\bibfnamefont {Y.-X.}\ \bibnamefont {Jiang}}, \bibinfo {author} {\bibfnamefont {M.}~\bibnamefont {Medarde}}, \bibinfo {author} {\bibfnamefont {X.}~\bibnamefont {Wu}}, \bibinfo {author} {\bibfnamefont {H.~C.}\ \bibnamefont {Lei}}, \bibinfo {author} {\bibfnamefont {J.}~\bibnamefont {Chang}},  \emph {et~al.},\ }\href@noop {} {\bibfield  {journal} {\bibinfo  {journal} {Nature}\ }\textbf {\bibinfo {volume} {602}},\ \bibinfo {pages} {245} (\bibinfo {year} {2022})}\BibitemShut {NoStop}%
\bibitem [{\citenamefont {Li}\ \emph {et~al.}(2022)\citenamefont {Li}, \citenamefont {Zhao}, \citenamefont {Ortiz}, \citenamefont {Park}, \citenamefont {Ye}, \citenamefont {Balents}, \citenamefont {Wang}, \citenamefont {Wilson},\ and\ \citenamefont {Zeljkovic}}]{li2022rotation}%
  \BibitemOpen
  \bibfield  {author} {\bibinfo {author} {\bibfnamefont {H.}~\bibnamefont {Li}}, \bibinfo {author} {\bibfnamefont {H.}~\bibnamefont {Zhao}}, \bibinfo {author} {\bibfnamefont {B.~R.}\ \bibnamefont {Ortiz}}, \bibinfo {author} {\bibfnamefont {T.}~\bibnamefont {Park}}, \bibinfo {author} {\bibfnamefont {M.}~\bibnamefont {Ye}}, \bibinfo {author} {\bibfnamefont {L.}~\bibnamefont {Balents}}, \bibinfo {author} {\bibfnamefont {Z.}~\bibnamefont {Wang}}, \bibinfo {author} {\bibfnamefont {S.~D.}\ \bibnamefont {Wilson}}, \ and\ \bibinfo {author} {\bibfnamefont {I.}~\bibnamefont {Zeljkovic}},\ }\href@noop {} {\bibfield  {journal} {\bibinfo  {journal} {Nature Physics}\ }\textbf {\bibinfo {volume} {18}},\ \bibinfo {pages} {265} (\bibinfo {year} {2022})}\BibitemShut {NoStop}%
\bibitem [{\citenamefont {Barman}\ \emph {et~al.}(2024)\citenamefont {Barman}, \citenamefont {Kim},\ and\ \citenamefont {Kim}}]{barman2024stacking}%
  \BibitemOpen
  \bibfield  {author} {\bibinfo {author} {\bibfnamefont {C.~K.}\ \bibnamefont {Barman}}, \bibinfo {author} {\bibfnamefont {S.-W.}\ \bibnamefont {Kim}}, \ and\ \bibinfo {author} {\bibfnamefont {Y.}~\bibnamefont {Kim}},\ }\href@noop {} {\bibfield  {journal} {\bibinfo  {journal} {Current Applied Physics}\ }\textbf {\bibinfo {volume} {68}},\ \bibinfo {pages} {31} (\bibinfo {year} {2024})}\BibitemShut {NoStop}%
\bibitem [{\citenamefont {Kim}\ \emph {et~al.}(2023)\citenamefont {Kim}, \citenamefont {Oh}, \citenamefont {Moon},\ and\ \citenamefont {Kim}}]{kim2023monolayer}%
  \BibitemOpen
  \bibfield  {author} {\bibinfo {author} {\bibfnamefont {S.-W.}\ \bibnamefont {Kim}}, \bibinfo {author} {\bibfnamefont {H.}~\bibnamefont {Oh}}, \bibinfo {author} {\bibfnamefont {E.-G.}\ \bibnamefont {Moon}}, \ and\ \bibinfo {author} {\bibfnamefont {Y.}~\bibnamefont {Kim}},\ }\href {https://www.nature.com/articles/s41467-023-36341-2} {\bibfield  {journal} {\bibinfo  {journal} {Nature Communications}\ }\textbf {\bibinfo {volume} {14}},\ \bibinfo {pages} {591} (\bibinfo {year} {2023})}\BibitemShut {NoStop}%
\bibitem [{\citenamefont {Song}\ \emph {et~al.}(2021)\citenamefont {Song}, \citenamefont {Ying}, \citenamefont {Chen}, \citenamefont {Han}, \citenamefont {Wu}, \citenamefont {Schnyder}, \citenamefont {Huang}, \citenamefont {Guo},\ and\ \citenamefont {Chen}}]{Songprl2021}%
  \BibitemOpen
  \bibfield  {author} {\bibinfo {author} {\bibfnamefont {Y.}~\bibnamefont {Song}}, \bibinfo {author} {\bibfnamefont {T.}~\bibnamefont {Ying}}, \bibinfo {author} {\bibfnamefont {X.}~\bibnamefont {Chen}}, \bibinfo {author} {\bibfnamefont {X.}~\bibnamefont {Han}}, \bibinfo {author} {\bibfnamefont {X.}~\bibnamefont {Wu}}, \bibinfo {author} {\bibfnamefont {A.~P.}\ \bibnamefont {Schnyder}}, \bibinfo {author} {\bibfnamefont {Y.}~\bibnamefont {Huang}}, \bibinfo {author} {\bibfnamefont {J.-g.}\ \bibnamefont {Guo}}, \ and\ \bibinfo {author} {\bibfnamefont {X.}~\bibnamefont {Chen}},\ }\href {\doibase 10.1103/PhysRevLett.127.237001} {\bibfield  {journal} {\bibinfo  {journal} {Phys. Rev. Lett.}\ }\textbf {\bibinfo {volume} {127}},\ \bibinfo {pages} {237001} (\bibinfo {year} {2021})}\BibitemShut {NoStop}%
\bibitem [{\citenamefont {Wang}\ \emph {et~al.}(2021)\citenamefont {Wang}, \citenamefont {Yu}, \citenamefont {Zhang}, \citenamefont {Liu}, \citenamefont {Li}, \citenamefont {Peng}, \citenamefont {Di}, \citenamefont {Jiang},\ and\ \citenamefont {Mu}}]{wang2021enhancement}%
  \BibitemOpen
  \bibfield  {author} {\bibinfo {author} {\bibfnamefont {T.}~\bibnamefont {Wang}}, \bibinfo {author} {\bibfnamefont {A.}~\bibnamefont {Yu}}, \bibinfo {author} {\bibfnamefont {H.}~\bibnamefont {Zhang}}, \bibinfo {author} {\bibfnamefont {Y.}~\bibnamefont {Liu}}, \bibinfo {author} {\bibfnamefont {W.}~\bibnamefont {Li}}, \bibinfo {author} {\bibfnamefont {W.}~\bibnamefont {Peng}}, \bibinfo {author} {\bibfnamefont {Z.}~\bibnamefont {Di}}, \bibinfo {author} {\bibfnamefont {D.}~\bibnamefont {Jiang}}, \ and\ \bibinfo {author} {\bibfnamefont {G.}~\bibnamefont {Mu}},\ }\href@noop {} {\bibfield  {journal} {\bibinfo  {journal} {arXiv preprint arXiv:2105.07732}\ } (\bibinfo {year} {2021})}\BibitemShut {NoStop}%
\bibitem [{\citenamefont {Zhang}\ \emph {et~al.}(2022)\citenamefont {Zhang}, \citenamefont {Wang}, \citenamefont {Tsang}, \citenamefont {Liu}, \citenamefont {Xie}, \citenamefont {Yu}, \citenamefont {Lai},\ and\ \citenamefont {Goh}}]{thinZhangPRB2022}%
  \BibitemOpen
  \bibfield  {author} {\bibinfo {author} {\bibfnamefont {W.}~\bibnamefont {Zhang}}, \bibinfo {author} {\bibfnamefont {L.}~\bibnamefont {Wang}}, \bibinfo {author} {\bibfnamefont {C.~W.}\ \bibnamefont {Tsang}}, \bibinfo {author} {\bibfnamefont {X.}~\bibnamefont {Liu}}, \bibinfo {author} {\bibfnamefont {J.}~\bibnamefont {Xie}}, \bibinfo {author} {\bibfnamefont {W.~C.}\ \bibnamefont {Yu}}, \bibinfo {author} {\bibfnamefont {K.~T.}\ \bibnamefont {Lai}}, \ and\ \bibinfo {author} {\bibfnamefont {S.~K.}\ \bibnamefont {Goh}},\ }\href {\doibase 10.1103/PhysRevB.106.195103} {\bibfield  {journal} {\bibinfo  {journal} {Phys. Rev. B}\ }\textbf {\bibinfo {volume} {106}},\ \bibinfo {pages} {195103} (\bibinfo {year} {2022})}\BibitemShut {NoStop}%
\bibitem [{\citenamefont {Zheng}\ \emph {et~al.}(2023)\citenamefont {Zheng}, \citenamefont {Tan}, \citenamefont {Chen}, \citenamefont {Wang}, \citenamefont {Zhu}, \citenamefont {Albarakati}, \citenamefont {Algarni}, \citenamefont {Partridge}, \citenamefont {Farrar}, \citenamefont {Zhou} \emph {et~al.}}]{zheng2023electrically}%
  \BibitemOpen
  \bibfield  {author} {\bibinfo {author} {\bibfnamefont {G.}~\bibnamefont {Zheng}}, \bibinfo {author} {\bibfnamefont {C.}~\bibnamefont {Tan}}, \bibinfo {author} {\bibfnamefont {Z.}~\bibnamefont {Chen}}, \bibinfo {author} {\bibfnamefont {M.}~\bibnamefont {Wang}}, \bibinfo {author} {\bibfnamefont {X.}~\bibnamefont {Zhu}}, \bibinfo {author} {\bibfnamefont {S.}~\bibnamefont {Albarakati}}, \bibinfo {author} {\bibfnamefont {M.}~\bibnamefont {Algarni}}, \bibinfo {author} {\bibfnamefont {J.}~\bibnamefont {Partridge}}, \bibinfo {author} {\bibfnamefont {L.}~\bibnamefont {Farrar}}, \bibinfo {author} {\bibfnamefont {J.}~\bibnamefont {Zhou}},  \emph {et~al.},\ }\href@noop {} {\bibfield  {journal} {\bibinfo  {journal} {Nature Communications}\ }\textbf {\bibinfo {volume} {14}},\ \bibinfo {pages} {678} (\bibinfo {year} {2023})}\BibitemShut {NoStop}%
\bibitem [{\citenamefont {Wang}\ and\ \citenamefont {Chubukov}(2014)}]{WangPRBfluc2014}%
  \BibitemOpen
  \bibfield  {author} {\bibinfo {author} {\bibfnamefont {Y.}~\bibnamefont {Wang}}\ and\ \bibinfo {author} {\bibfnamefont {A.}~\bibnamefont {Chubukov}},\ }\href {\doibase 10.1103/PhysRevB.90.035149} {\bibfield  {journal} {\bibinfo  {journal} {Phys. Rev. B}\ }\textbf {\bibinfo {volume} {90}},\ \bibinfo {pages} {035149} (\bibinfo {year} {2014})}\BibitemShut {NoStop}%
\bibitem [{\citenamefont {Sarkar}\ \emph {et~al.}(2019)\citenamefont {Sarkar}, \citenamefont {Chakraborty},\ and\ \citenamefont {P\'epin}}]{sarkarloopcurrentprb2019}%
  \BibitemOpen
  \bibfield  {author} {\bibinfo {author} {\bibfnamefont {S.}~\bibnamefont {Sarkar}}, \bibinfo {author} {\bibfnamefont {D.}~\bibnamefont {Chakraborty}}, \ and\ \bibinfo {author} {\bibfnamefont {C.}~\bibnamefont {P\'epin}},\ }\href {\doibase 10.1103/PhysRevB.100.214519} {\bibfield  {journal} {\bibinfo  {journal} {Phys. Rev. B}\ }\textbf {\bibinfo {volume} {100}},\ \bibinfo {pages} {214519} (\bibinfo {year} {2019})}\BibitemShut {NoStop}%
\bibitem [{\citenamefont {Sachdev}(2018)}]{sachdev2018topological}%
  \BibitemOpen
  \bibfield  {author} {\bibinfo {author} {\bibfnamefont {S.}~\bibnamefont {Sachdev}},\ }\href@noop {} {\bibfield  {journal} {\bibinfo  {journal} {Reports on Progress in Physics}\ }\textbf {\bibinfo {volume} {82}},\ \bibinfo {pages} {014001} (\bibinfo {year} {2018})}\BibitemShut {NoStop}%
\bibitem [{\citenamefont {B\"oker}\ \emph {et~al.}(2017)\citenamefont {B\"oker}, \citenamefont {Volkov}, \citenamefont {Efetov},\ and\ \citenamefont {Eremin}}]{volkov}%
  \BibitemOpen
  \bibfield  {author} {\bibinfo {author} {\bibfnamefont {J.}~\bibnamefont {B\"oker}}, \bibinfo {author} {\bibfnamefont {P.~A.}\ \bibnamefont {Volkov}}, \bibinfo {author} {\bibfnamefont {K.}~\bibnamefont {Efetov}}, \ and\ \bibinfo {author} {\bibfnamefont {I.}~\bibnamefont {Eremin}},\ }\href@noop {} {\bibfield  {journal} {\bibinfo  {journal} {Phys. Rev. B}\ }\textbf {\bibinfo {volume} {96}},\ \bibinfo {pages} {04517} (\bibinfo {year} {2017})}\BibitemShut {NoStop}%
\bibitem [{\citenamefont {Timoshuk}\ and\ \citenamefont {Babaev}()}]{babaev}%
  \BibitemOpen
  \bibfield  {author} {\bibinfo {author} {\bibfnamefont {I.}~\bibnamefont {Timoshuk}}\ and\ \bibinfo {author} {\bibfnamefont {E.}~\bibnamefont {Babaev}},\ }\href@noop {} {\bibfield  {journal} {\bibinfo  {journal} {arXiv}\ }\textbf {\bibinfo {volume} {2407.20132}}}\BibitemShut {NoStop}%
\bibitem [{\citenamefont {Park}\ \emph {et~al.}(2021)\citenamefont {Park}, \citenamefont {Ye},\ and\ \citenamefont {Balents}}]{Park}%
  \BibitemOpen
  \bibfield  {author} {\bibinfo {author} {\bibfnamefont {T.}~\bibnamefont {Park}}, \bibinfo {author} {\bibfnamefont {M.}~\bibnamefont {Ye}}, \ and\ \bibinfo {author} {\bibfnamefont {L.}~\bibnamefont {Balents}},\ }\href {\doibase 10.1103/PhysRevB.104.035142} {\bibfield  {journal} {\bibinfo  {journal} {Phys. Rev. B}\ }\textbf {\bibinfo {volume} {104}},\ \bibinfo {pages} {035142} (\bibinfo {year} {2021})}\BibitemShut {NoStop}%
\bibitem [{\citenamefont {Anber}\ \emph {et~al.}(2012)\citenamefont {Anber}, \citenamefont {Poppitz},\ and\ \citenamefont {\"Unsal}}]{anber20122d}%
  \BibitemOpen
  \bibfield  {author} {\bibinfo {author} {\bibfnamefont {M.~M.}\ \bibnamefont {Anber}}, \bibinfo {author} {\bibfnamefont {E.}~\bibnamefont {Poppitz}}, \ and\ \bibinfo {author} {\bibfnamefont {E.~E.}\ \bibnamefont {\"Unsal}},\ }\href@noop {} {\bibfield  {journal} {\bibinfo  {journal} {J. High Energy Physics}\ }\textbf {\bibinfo {volume} {2012}},\ \bibinfo {pages} {1} (\bibinfo {year} {2012})}\BibitemShut {NoStop}%
\bibitem [{\citenamefont {Anber}\ \emph {et~al.}(2013)\citenamefont {Anber}, \citenamefont {Collier},\ and\ \citenamefont {Poppitz}}]{anber20133}%
  \BibitemOpen
  \bibfield  {author} {\bibinfo {author} {\bibfnamefont {M.~M.}\ \bibnamefont {Anber}}, \bibinfo {author} {\bibfnamefont {S.}~\bibnamefont {Collier}}, \ and\ \bibinfo {author} {\bibfnamefont {E.}~\bibnamefont {Poppitz}},\ }\href@noop {} {\bibfield  {journal} {\bibinfo  {journal} {J. High Energy Physics}\ }\textbf {\bibinfo {volume} {2013}},\ \bibinfo {pages} {1} (\bibinfo {year} {2013})}\BibitemShut {NoStop}%
\bibitem [{Note1()}]{Note1}%
  \BibitemOpen
  \bibinfo {note} {Instead of simulating the Boltzmann factor $e^{-\protect \mathcal {F}/T}$, one can equivalently evaluate the factor $e^{-\protect \mathcal {H}/T}$ from Eqs.~\protect \eqref {ham}.}\BibitemShut {Stop}%
\end{thebibliography}%
\end{document}